# Quantifying human gray matter microstructure using NEXI and 300 mT/m gradients

*Quentin Uhl[1,2], Tommaso Pavan[1,2], Malwina Molendowska[3], Derek K. Jones[3], Marco Palombo[3,4,\*], Ileana Jelescu[1,2,\*]*

[1]Department of Radiology, Lausanne University Hospital (CHUV), Lausanne, Switzerland, [2]School of Biology and Medicine, University of Lausanne, Lausanne, Switzerland, [3]Cardiff University Brain Research Imaging Centre (CUBRIC), Cardiff University, Cardiff, United Kingdom, [4]School of Computer Science and Informatics, Cardiff University, Cardiff, UK, *joint last authorship

## Abstract

Biophysical models of diffusion tailored to quantify gray matter microstructure are gathering increasing interest. The two-compartment Neurite EXchange Imaging (NEXI) model has been proposed recently to account for neurites, extra-cellular space and exchange across the cell membrane. NEXI parameter estimation requires multi-shell multi-diffusion time data and has so far only been implemented experimentally on animal data collected on a preclinical MRI set-up. In this work, the first ever translation of NEXI to the human cortex in vivo was achieved using a 3T Connectom MRI system with 300 mT/m gradients, that enables the acquisition of a broad range of *b*-values (0 – 7.5 ms/µm²) with a window of diffusion times (20 – 49 ms) suitable for the expected characteristic exchange times (10 – 50 ms). Microstructure estimates of four model variants: NEXI, $NEXI_{dot}$ (its extension with the addition of a dot compartment) and their respective versions that correct for the Rician noise floor ($NEXI_{RM}$ and $NEXI_{dot,RM}$) that particularly impacts high b-value signal, were compared. The reliability of estimates in each model variant was evaluated in synthetic and human in vivo data. In the latter, the intra-subject (scan-rescan) vs between-subjects variability of microstructure estimates were compared in the cortex. The better performance of $NEXI_{RM}$ highlights the importance of correcting for Rician bias in the NEXI model to obtain accurate estimates of microstructure parameters in the human cortex, and the sensitivity of the NEXI framework to individual differences in cortical microstructure. This groundbreaking application of NEXI in humans marks a pivotal moment, unlocking new avenues for studying neurodevelopment, ageing, and various neurodegenerative disorders.

## 1. Introduction

Quantifying microstructure features of the human cortex in vivo has the potential to significantly improve our understanding and management of neurological and psychiatric diseases, which are associated with cognitive, motor, and behavioral deficits (Illán-Gala et al., 2022; Nürnberger et al., 2017; Spotorno et al., 2022; Voldsbekk et al., 2022). Early diagnosis and effective treatment of these diseases remain a challenge, as their pathophysiology is not fully understood. Identifying the associated changes in the cortex microstructure could lead to a better understanding of the disease progression, earlier diagnoses and access to treatment, and help develop targeted therapies.

Diffusion-weighted magnetic resonance imaging (dMRI) can provide such an insight into the microstructure of the brain, by exploiting the sensitivity of the signal to the motion of water molecules within tissue. In particular, biophysical modeling of the dMRI signal aims to characterize the tissue microstructure by fitting an analytical model of the tissue described by its most relevant geometric and diffusion features (Alexander et al., 2019; Jelescu et al., 2020; Novikov et al., 2019, 2018; Stanisz et al., 1997) to the measured signals.

There is already a wide variety of biophysical models of white matter, based on what is now commonly referred to as the "Standard Model" (Novikov et al., 2019) of non-exchanging compartments within which the diffusion displacement profile is Gaussian. However, recent studies indicate that the Standard Model does not hold in gray matter. At high b-values, the deviation of the directionally-averaged signal in gray matter from the impermeable stick power-law $\bar{S} \propto b^{-1/2}$ (McKinnon et al., 2017; Veraart et al., 2016a) prompted the hypotheses that other features such as the cell body or 'soma' (Palombo et al., 2020, 2018), inter-compartment exchange (Jelescu et al., 2022; Olesen et al., 2022; Veraart et al., 2018) and non-Gaussian diffusion within a compartment resulting from structural disorder (Henriques et al., 2019; Lee et al., 2020) should be accounted for. Indeed, in the cortex, most neurites are unmyelinated, so that the exchange of water between the intracellular and extracellular compartments may be significant for diffusion times that are longer than 20 ms (typical of the minimal diffusion time achievable on human MRI scanners). Additionally, the assumption of Gaussian diffusion within a given compartment may not hold in the presence of irregularities on length scales that are similar to the diffusion length, such as dendritic spines and neurite beading. Furthermore, the volume occupied by soma, in the gray matter is approximately 10-20%, but negligible in white matter and therefore not currently included in white matter models.

As an extension of the Standard Model, the Soma And Neurite Density Imaging (SANDI) model (Palombo et al., 2020), incorporated the soma size and signal fraction in addition to neurite signal fraction, thereby enabling their joint estimation. However, as it does not account for inter-compartment exchange, the SANDI model is currently only applicable to data acquired within diffusion times shorter than 20 ms, for which the assumption of impermeable compartments is valid (Jelescu et al., 2020). As noted above, such diffusion times can only be achieved for very high b-values (up to 10 ms/µm²), on systems with ultra-strong gradients, such as preclinical scanners or human scanners with dedicated gradient sets (such as the Connectom scanner, 300 mT/m gradient amplitude) (Huang et al., 2021; Jones et al., 2018; Setsompop et al., 2013).

The Neurite Exchange Imaging (NEXI) model (Jelescu et al., 2022) – proposed in parallel by (Olesen et al., 2022) as SMEX (Standard Model with EXchange) – was introduced recently to recognize and quantify water exchange across the neurite membrane. As such, NEXI is applicable on clinical grade scanners because it does not necessarily require short diffusion times. NEXI models the neurites as a collection of randomly-oriented sticks – occupying a relative signal fraction $f$ – where the intra-neurite diffusion is uniaxial with diffusivity $D_{i,\parallel}$. Moreover, given the quasi-uniform orientation-distribution of neurites in gray matter, the extra-neurite compartment is considered to be Gaussian isotropic with characteristic diffusivity $D_e$. The two compartments exchange with a characteristic time $t_{ex}$. NEXI models the total orientation-averaged signal as the sum of these two exchanging compartments. They are assumed to have the same transverse relaxation time, or $T_2$. The soma are not explicitly modeled and the signal contribution arising from this compartment is most likely pooled with the signal contribution from the extra-cellular space in NEXI (Jelescu et al., 2022). Importantly, the experimental observation of decreasing signal with increasing diffusion times supports exchange as a dominant contributor to signal features over a soma compartment with restricted diffusion (Jelescu et al., 2022; Olesen et al., 2022), although accounting for soma improves the fit of the signal tail (highest b-values). Thus, if the available diffusion MRI data do not allow fitting a model with enough parameters to account for both exchange and soma, modeling exchange while neglecting soma can be justified for diffusion times $t_d$ longer than 20ms. On the other hand, an extension of SMEX which also models the soma as a separate compartment (SANDIX – SANDI with eXchange) has been proposed and applied to ex-vivo preclinical data (Olesen et al., 2022). The stability of fitting such a large number of model parameters on human in-vivo data remains to be established.

The NEXI signal equation is a spherical mean of the kernel $\mathcal{K}$, the anisotropic extension of the Kärger model of two well-mixed exchanging compartments in a barrier-limited regime (Fieremans et al., 2010; Jelescu et al., 2022; Kärger, 1985):

$$\bar{S}_{NEXI}(\boldsymbol{p}; q, t_d) = \int_0^1 \mathcal{K}(q, \boldsymbol{g}, t_d; \boldsymbol{p}, \boldsymbol{n}) d(\boldsymbol{g}.\boldsymbol{n})^2 \qquad (1)$$

where $\boldsymbol{p} = [t_{ex}, D_{i,\parallel}, D_e, f]$ are the microstructure parameters to fit, **n** are the neurite orientations, $q$ is the wave vector along direction **g**.

The assumption of the barrier-limited regime is supported if the characteristic time $t_c$ to reach the long-time diffusion limit in each compartment is shorter than the characteristic exchange time between compartments. In the case of infinitely long cylinders modeling the neurites, the radial plane is relevant for exchange across the membrane. In the case of neurites with a diameter d ~ 1 µm, the characteristic time in the intra-neurite space $t_{c,i} = \frac{d^2}{2D_i} \approx 0.25\ ms$ and extra-neurite space $t_{c,e} = \frac{d^2}{2D_e}\sqrt{\frac{\pi}{f}} \approx 2.3\ ms$ at most – assuming the lower bound of $f$~0.3 (Fieremans et al., 2010). Both timescales are shorter than the exchange time reported in previous studies $t_{ex} > 5$ ms. We note that the Kärger model assumption implies diffusion should be time-independent, while some time-dependence has been reported in a previous in vivo study of the human cortex (Lee et al., 2020), D(t) was weak and the long-time limit was reached for $t_d$ > 20 ms, which agrees with the experimental setting in the present study.

The aim of this study was to evaluate the feasibility and value of using the NEXI model and some of its variants for quantifying microstructural parameters in the human cortex in vivo.

To achieve this, we acquired multi-shell multi-diffusion time dMRI data in healthy human volunteers on a Connectom MRI system equipped with very strong 300 mT/m gradients. The Connectom scanners are an important steppingstone in terms of hardware capabilities between preclinical MRI systems (with gradients >600 mT/m) and clinical MRI systems (with gradients ≤ 80 mT/m). They provide the opportunity of an initial translation of NEXI in human subjects by enabling the acquisition of the necessary broad range of b-values (0 – 7.5 ms/µm²) at diffusion times 20 – 49 ms, that are short enough to capture exchange processes with expected $t_{ex}$ = 10 - 50 ms, as previously reported for brain cortex *in vivo* (Jelescu et al., 2022; Lee et al., 2020).

Here, we compared NEXI-derived estimates in the human cortex to those obtained from its three-compartment variant, allowing for an extra 'dot' compartment, filled with stationary water. This NEXI extension, referred to here as NEXI$_{dot}$, has been proposed previously (Olesen et al., 2022) to explain the non-zero signal asymptote at high b-value ex vivo. In the cerebellum, the presence of such a compartment has been shown in vivo (Tax et al., 2020), but its existence in the cortex remains unclear. This compartment's stationary water signal does not decay with diffusion-weighting, thus yielding the NEXI$_{dot}$ signal attenuation equation:

$$\bar{S}_{NEXI_{dot}}(\boldsymbol{p}; q, t_d) = (1 - f_{dot}).\bar{S}_{NEXI}(\boldsymbol{p}; q, t_d) + f_{dot} \qquad (2)$$

where $f_{dot}$ is the stationary water fraction.

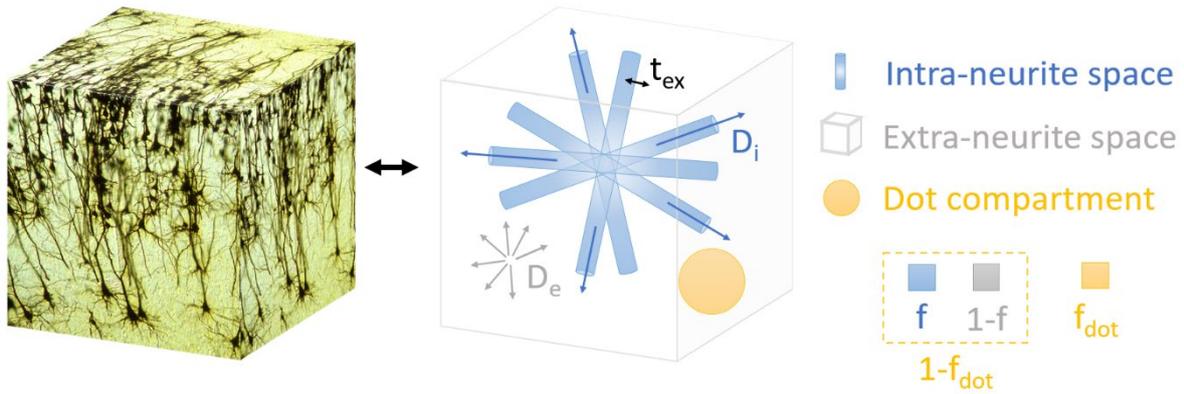

**Figure 1.** Sketch of relevant features and parameters in the NEXI$_{dot}$ model, the three-compartment variant of NEXI. The latter can be obtained by removing the dot compartment (in yellow) from the sketch.

At high b-values, high spatial resolution and moderate field strength, the diffusion-weighted signal magnitude is heavily affected by the Rician noise floor. The effect of this noise floor can be accounted for by considering the expectation value of the signal $\bar{S}_{NEXI}(\boldsymbol{p}; q, t_d)$ given the normalized Rician noise level $\sigma = \frac{\sigma_{Signal}}{S_{b=0}}$. The NEXI signal equation corrected for the Rician Mean (RM) is:

$$\bar{S}_{NEXI_{RM}}(\boldsymbol{p}; q, t_d, \sigma) = \sqrt{\frac{\pi}{2}} \cdot \sigma \cdot L_{1/2}\left(-\frac{1}{2}\left(\frac{\bar{S}_{NEXI}(\boldsymbol{p}; q, t_d)}{\sigma}\right)^2\right) \quad (3)$$

where $L_{1/2}(x) = {}_1F_1\left(-\frac{1}{2}, 1, x\right)$ is the generalized Laguerre polynomial, expressed in terms of the confluent hypergeometric function of the first kind.

Similarly, the signal equation for the NEXI$_{dot,RM}$ model is:

$$\bar{S}_{NEXI_{dot,RM}}(\boldsymbol{p}; q, t_d, \sigma) = \sqrt{\frac{\pi}{2}} \cdot \sigma \cdot L_{1/2}\left(-\frac{1}{2}\left(\frac{\bar{S}_{NEXI_{dot}}(\boldsymbol{p}; q, t_d)}{\sigma}\right)^2\right) \quad (4)$$

We therefore compared NEXI and NEXI$_{dot}$ estimates to their respective RM-corrected counterparts.

Furthermore, we compared the estimates of $t_{ex}$ from the different model variants with the one from the Kärger model time-dependent kurtosis (Els Fieremans et al., 2010; Jelescu et al., 2022; Jensen and Helpern, 2010):

$$K_{KM}(t_d) = 2\frac{t_{ex}}{t_d}\left[1 - \frac{t_{ex}}{t_d}\left(1 - e^{\frac{t_d}{t_{ex}}}\right)\right] \quad (5)$$

Finally, we estimated the repeatability and sensitivity of NEXI cortex microstructure estimates by comparing their intra-subject (scan-rescan) to inter-subject variability. Parameter spatial distribution across different brain regions was also evaluated in comparison with known distribution maps from postmortem histological staining.

## 2. Methods

## 2.1 Experimental

### 2.1.1 Participants

The study was approved by the School of Psychology Ethics Committee Cardiff University. Written informed consent was obtained from all participants. Data were acquired in four healthy adults (Age: 30.5 +/- 3.8 years; 2 M / 2 F). Three participants were rescanned two days after the first scan.

### 2.1.2 Data acquisition

All data were acquired on a Connectom MRI scanner, a modified 3T MAGNETOM Skyra system fitted with a gradient coil capable of 300 mT/m (Siemens Healthcare, Erlangen, Germany). An anatomical reference was acquired using an MP-RAGE sequence (1-mm isotropic resolution, FOV = 256 x 256 mm$^2$, 192 slices, TI/TR = 857/2300 ms). Diffusion-weighted images were acquired using a Pulsed Gradient Spin Echo Echo-Planar Imaging (PGSE EPI) sequence with b-values of 1 (13 directions), 2.5 (25 dir.), 4 (25 dir.), 6 (32 dir.) and 7.5 ms/µm² (65 dir.), at each of four diffusion times Δ = 20, 29, 39 and 49 ms, in addition to 15 b = 0 ms/µm² images per Δ. Other parameters were fixed: δ = 9 ms, TE/TR = 76 ms/3.7 s, FOV = 216 x 216 mm$^2$, matrix: 120x120, 66 slices, 1.8-mm isotropic resolution, partial Fourier = 0.75, GRAPPA = 2, multiband = 2. The total dMRI scan time was 45 min.

### 2.1.3 Data preprocessing

While each diffusion time was acquired in a separate scan, all multi-shell multi-diffusion time data (N = 700 volumes) were pooled together for pre-processing. Pre-processing included Marchenko-Pastur principal component analysis (MP-PCA) magnitude denoising (Veraart et al., 2016b), Gibbs ringing correction (Kellner et al., 2016), distortion and eddy current correction (Andersson and Sotiropoulos, 2016). A separate MP-PCA denoising of $b$ = 0 and $b$ = 1 ms/µm² images (N = 112 volumes) was used to extract an unbiased noisemap, σ, from high SNR data, to be used in the Rician mean correction (Eq. 3-4). For NEXI, data were averaged over directions (powder-average, using the arithmetic mean) and normalized by the mean value of the $b$ = 0 ms/µm² volumes.

### 2.1.4 Time-dependent kurtosis

DKI fitting (Jensen et al., 2005) was performed using a weighted linear least squares algorithm implemented in Matlab (Veraart et al., 2013) to extract Mean Diffusivity (MD) and Mean Kurtosis (MK) for each diffusion time using $b$-values up to 2.5 ms/µm². $K_{KM}(t)$ (Eq. 5) was then fit to MK to yield an alternative estimation of $t_{ex}$.

### 2.1.5 ROI parcellation.

Grey matter region of interests (ROIs) from the Desikan-Killiany-Tourville (DKT) atlas (Klein and Tourville, 2012) were segmented on the anatomical MPRAGE image using FastSurfer (Henschel et al., 2020) and transformed into diffusion native space using linear registration of distortion-corrected b = 0 ms/µm² images to MPRAGE images. The cortical ribbon was segmented by merging the gray matter ROIs obtained with the DKT atlas.

## 2.2 Simulations

Synthetic NEXI signals were generated using Eq. 1 and the same diffusion times and $b$-values as the experimental acquisition. The ground truth parameters of each signal were randomly chosen within the following bounds with uniform probability distribution: [1 - 150] ms for $t_{ex}$, [0.1 - 3.5] µm²/ms for the two diffusivities and [0.1 - 0.9] the fraction $f$, with the constraint that $D_i > D_e$ (Dhital et al., 2019; Howard et al., 2022; Kunz et al., 2018). Twenty Rician noise realizations were generated for each ground truth, assuming SNR = 34 at b = 0 ms/µm² (as estimated from our in vivo data), and then averaged to mimic powder-averaging of magnitude images, which increases the SNR but does not lower the Rician floor. A dataset of 10,000 ground truth combinations was generated in this way.

**2.3 Comparison between NEXI model variants**

The four NEXI model variants (Eq. 1-4) were fit to the synthetic and experimental data by Nonlinear Least Squares (NLS) using the L-BFGS-B algorithm and minimize function from the package scipy.optimize (Virtanen et al., 2020), with a tolerance of 1e-14. The bounds specified for the optimization were the same as those previously described for the simulations. For the models with a dot compartment, $f_{dot}$ bounds were [0.0001 0.3]. For the models with Rician mean correction, σ was fixed to the noise level estimated in 2.1.3 for experimental data, and to the noise level set in the simulations for synthetic data. To assess the impact of a misestimation of σ in MP-PCA on the performance of $NEXI_{RM}$, σ was also fixed to a value overestimated by 10%, 20% and 50% of the actual noise level set in the simulations. The metric used for the optimization was the Mean Square Error (MSE) of the estimated signals against the measured or simulated signals. An initial grid search was applied before the NLS to find an optimal starting point.

**2.3.1 Performance in synthetic data**

The comparison of the model performance was based on the Median Absolute Error (MedAE) and Root Mean Square Error (RMSE) between ground truth and estimation of each model, on the four parameters of interest. These two metrics were chosen to observe both the real performance of the model and the variance of this performance. The MedAE is more robust to outliers and thus more representative of the performance of the model. The RMSE is complementary to this measure, providing information on both bias and variance of the model, but remains very sensitive to outliers.

**2.3.2 Performance in experimental data**

To compare the fit of the four models on our experimental data, one of the criteria used was the corrected Akaike Information Criterion (AICc) (Akaike, 1973).
Furthermore, since both the dot compartment and the Rician noise floor account for the diffusion signal not decaying asymptotically to zero, the dot compartment estimation $f_{dot}$ in $NEXI_{dot}$ was compared to the Rician floor derived from the noise standard deviation in each ROI, estimated using MP-PCA and used as an input to $NEXI_{RM}$.

**2.3.3 Repeatability and brain region-specific patterns**

Intra-subject vs inter-subject variability was assessed on average GM median ROI estimates obtained by the $NEXI_{RM}$ model using Bland-Altman plots (Altman and Bland, 1983).

The spatial distribution of GM microstructure features quantified using $NEXI_{RM}$ was also examined using inflated brain surfaces obtained using Connectome Workbench (Marcus et al., 2011) and compared to distribution patterns of neurite density and myelination from the Glasser MRI atlas (Glasser et al., 2022).

# 3. Results

## 3.1 Simulations

Estimation errors on synthetic data (**Table 1**) show the NEXI$_{RM}$ model yields $t_{ex}$ estimates with a 25% lower MedAE and 20% lower RMSE compared to all the other model variants. The estimates of neurite fraction $f$ and extracellular diffusivity $D_e$ are also substantially improved using the NEXI$_{RM}$ model, lowering the MedAE by at least 15% and 20%, respectively.

A.

| MedAE | $t_{ex}$ (ms) | $D_i$ (µm²/ms) | $D_e$ (µm²/ms) | $f$ |
|---|---|---|---|---|
| NEXI | 30.1 | 0.452 | 0.076 | 0.052 |
| NEXI$_{dot}$ | 31.2 | 0.507 | 0.092 | 0.068 |
| NEXI$_{RM}$ | **22.2** | **0.393** | **0.060** | **0.044** |
| NEXI$_{dot,RM}$ | 30.2 | 0.466 | 0.089 | 0.063 |

B.

| RMSE | $t_{ex}$ (ms) | $D_i$ (µm²/ms) | $D_e$ (µm²/ms) | $f$ |
|---|---|---|---|---|
| NEXI | 51.1 | 1.101 | 0.327 | 0.133 |
| NEXI$_{dot}$ | 55.7 | 0.982 | 0.447 | 0.154 |
| NEXI$_{RM}$ | **41.5** | 0.916 | **0.320** | **0.121** |
| NEXI$_{dot,RM}$ | 55.0 | **0.898** | 0.446 | 0.155 |

**Table 1.** MedAE (A.) and RMSE (B.) of the different model variants on the synthetic dataset. Note the synthetic data spanned broad parameter ranges of ground truths, thus these summary statistics are only partially informative. Due to the bounded estimator, the MedAE is a less biased indicator than the RMSE.

Due to the broad parameter ranges spanned by the synthetic data ground truths, a binned representation of estimation error is more informative (**Figure 2**). For parameters with the highest estimation uncertainty, $t_{ex}$ and $D_i$, the upper and lower bounds on the estimation yielded very asymmetric distributions for bins with ground truth values near those bounds (e.g. for $t_{ex}$ target ~140 ms or $D_i$ target ~ 3.0 µm²/ms). Thus, the metrics in **Table 1** are affected by these constraints, mainly the RMSE which is more sensitive to extreme values.

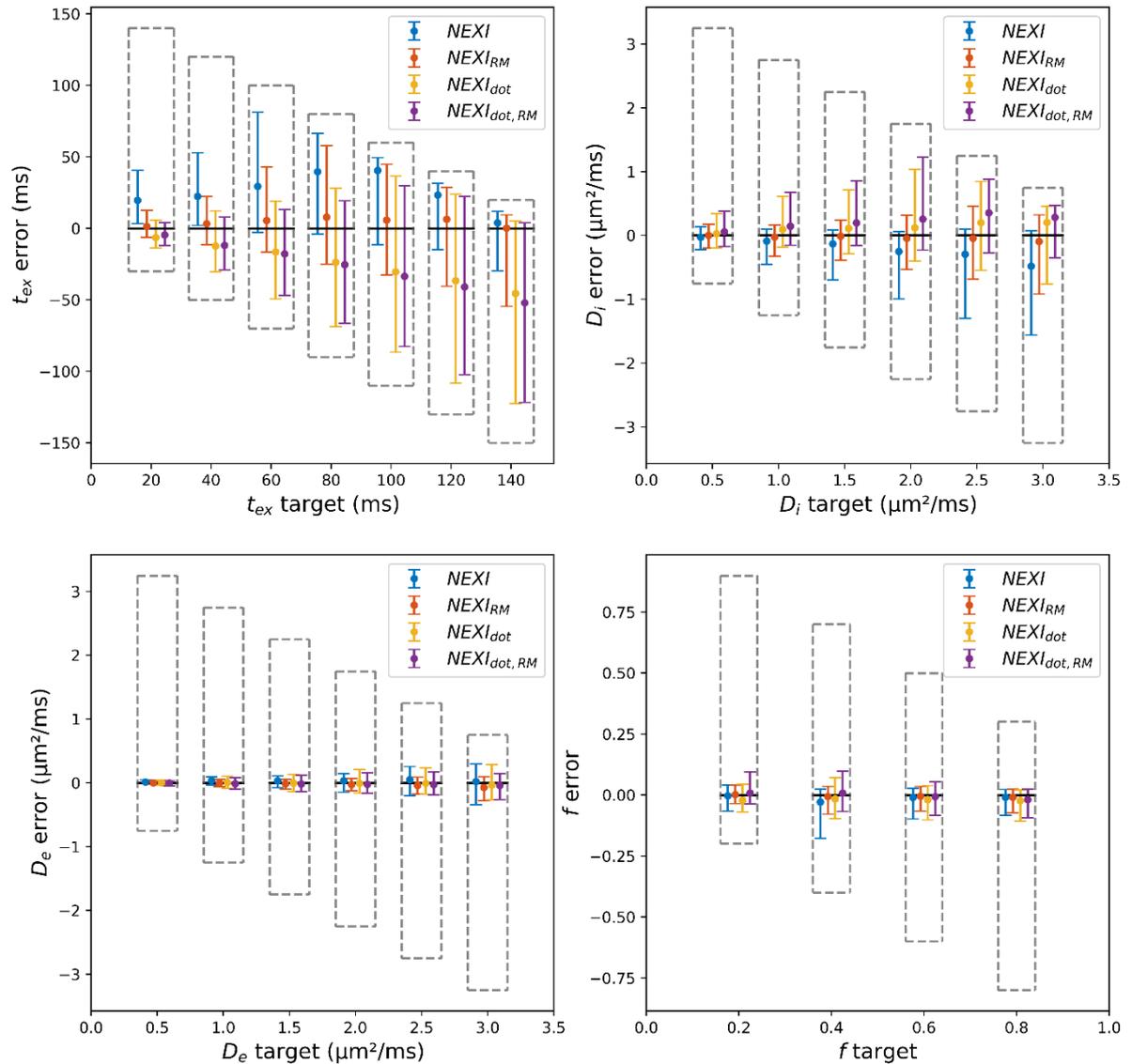

**Figure 2.** Boxplots (median and interquartile range) of parameter estimates by each of the four model variants on synthetic datasets with random Rician noise (σ = 0.03). The error is defined as the difference between the estimation and the target value. The upper and lower limits of the grey dashed box represent the maximum and minimum error of the estimator, in the given bin, due to lower and upper bounds in the NLS algorithm.

The boxplots (**Figure 2**) confirm that, assuming a ground truth of two exchanging compartments and Rician noise, NEXI$_{RM}$ yields the best accuracy and precision across model parameters and their biologically plausible ranges of values. The neurite fraction $f$ and extra-cellular diffusivity $D_e$ estimates benefit from good to excellent accuracy and precision with any model variant. For the two parameters with higher bias and uncertainty ($t_{ex}$ and $D_i$): The accuracy on $t_{ex}$ and $D_i$ was markedly reduced using NEXI, NEXI$_{dot}$ and NEXI$_{dot,RM}$ as compared to NEXI$_{RM}$, as well as the precision on $t_{ex}$ using NEXI$_{dot}$ and NEXI$_{dot,RM}$.

**A.**

| MedAE of NEXI$_{RM}$ using: | $t_{ex}$ (ms) | $D_i$ (µm²/ms) | $D_e$ (µm²/ms) | $f$ |
|---|---|---|---|---|
| Ground truth σ | 22.2 | 0.393 | 0.060 | 0.044 |
| 110% σ | 22.8 | 0.389 | 0.065 | 0.045 |
| 120% σ | 24.6 | 0.389 | 0.073 | 0.047 |
| 150% σ | 31.1 | 0.413 | 0.099 | 0.055 |

B.

| RMSE of NEXI$_{RM}$ using: | $t_{ex}$ (ms) | $D_i$ (μm²/ms) | $D_e$ (μm²/ms) | $f$ |
|---|---|---|---|---|
| Ground truth σ | 41.5 | 0.916 | 0.320 | 0.121 |
| 110% σ | 42.5 | 0.896 | 0.355 | 0.128 |
| 120% σ | 44.9 | 0.895 | 0.386 | 0.138 |
| 150% σ | 52.2 | 0.940 | 0.476 | 0.157 |

**Table 2.** MedAE (A.) and RMSE (B.) of NEXI$_{RM}$ using the true and overestimated σ, on a synthetic dataset with random Rician noise (true σ = 0.03). Note the synthetic data spanned broad parameter ranges of ground truths, thus these summary statistics are only partially informative. Due to the bounded estimator, the MedAE is a less biased indicator than the RMSE.

For a 50% overestimation of σ, the NEXI$_{RM}$ errors are comparable to those of the other models (**Table 2** and **Supplementary Figure S1**). This indicates that some error in the σ estimation from MP-PCA can be tolerated within the NEXI$_{RM}$ model. Releasing σ as a free model parameter in NEXI$_{RM}$ yielded either similar values to MP-PCA, or a convergence of σ to zero and poorer AICc (data not shown).

Since the synthetic data were generated assuming a model of two exchanging compartments, it is expected that NEXI$_{RM}$ variants perform better than NEXI$_{dot}$ variants. However, the simulations underline that failing to account for the Rician floor in the NEXI fit, when Rician noise is present in the data, results in a drastic deterioration of the quality of estimates (NEXI vs NEXI$_{RM}$). They also reveal that the dot compartment fails to mitigate the error due to Rician noise. When introducing a dot compartment in the model while it is not present in the data results in a deterioration of estimates for all other model parameters, in particular for the exchange time (NEXI$_{dot}$ and NEXI$_{dot,RM}$ vs NEXI$_{RM}$).

### 3.2 Experimental

Based on the DKT parcellation, median values across GM ROIs for each of the model variants are presented in Table 3. The four model variants give very different exchange time estimates. Notably, $t_{ex}$ estimates are ordered as NEXI > NEXI$_{RM}$ > NEXI$_{dot}$. All these estimates are also much longer than 3-5 ms, as reported using NEXI$_{dot,RM}$ (though the latter was comparable to NEXI$_{dot}$ in the simulations) and previously in *ex vivo* data (Jelescu and Uhl, 2022; Olesen et al., 2022). The extra-neurite diffusivity estimates are comparable across methods. Three of the four models give an intra-neurite diffusivity very close to the upper limit, indicating that the model often hit the bounds, and it may be missing a component to explain experimental data well. The first three methods seem to agree for an average $f$ around 0.35 while NEXI$_{dot,RM}$ places it higher, at 0.47.

When comparing between models with and between models without Rician mean correction, NEXI$_{dot}$ has a better corrected AICc than NEXI, but the opposite happens when we add the Rician correction, NEXI$_{RM}$ outperforms NEXI$_{dot,RM}$.

| | t$_{ex}$ (ms) | D$_i$ (μm²/ms) | D$_e$ (μm²/ms) | f | f$_{dot}$ | AICc |
|---|---|---|---|---|---|---|
| **NEXI** | 103.9 [100.3, 107.5] | 2.79 [2.71, 2.88] | 0.95 [0.94, 0.96] | 0.32 [0.318, 0.325] | - | **42.49** ± 5.62 |
| **NEXI$_{dot}$** | 14.3 [12.2, 16.3] | 3.36 [3.32, 3.40] | 1.00 [0.99, 1.01] | 0.36 [0.35, 0.37] | 0.03 [0.033, 0.037] | **38.08** ± 6.18 |
| **NEXI$_{RM}$** | 42.3 [40.0, 44.7] | 3.35 [3.32, 3.38] | 0.92 [0.91, 0.93] | 0.38 [0.379, 0.389] | - | **45.45** ± 5.69 |
| **NEXI$_{dot,\ RM}$** | 2.90 [2.71, 3.09] | 3.36 [3.34, 3.39] | 1.03 [1.01, 1.04] | 0.47 [0.47, 0.48] | 0.01 [0.009, 0.010] | **48.24** ± 6.19 |

**Table 3.** Mean estimates and 95% confidence intervals of the median in every ROI of the DKT atlas using NEXI, NEXI$_{dot}$, Corrected for Rice Mean (RM) or not. The last column shows the mean corrected Akaike Information Criterion (AICc) for each model; lower AICc indicates a better fit.

The mean fitted powder-average signal in the whole cortical ribbon by the four model variants is shown in **Figure 3**. The quality of fit shows that at high b-value and high diffusion time, NEXI performs poorly compared to the other models. However, there is limited agreement between the mean signal and all the models mean fitting curves at high b-value. This is due to the trade-off of fitting the signal across the entire b-value range (Supplementary **Figure S2**).

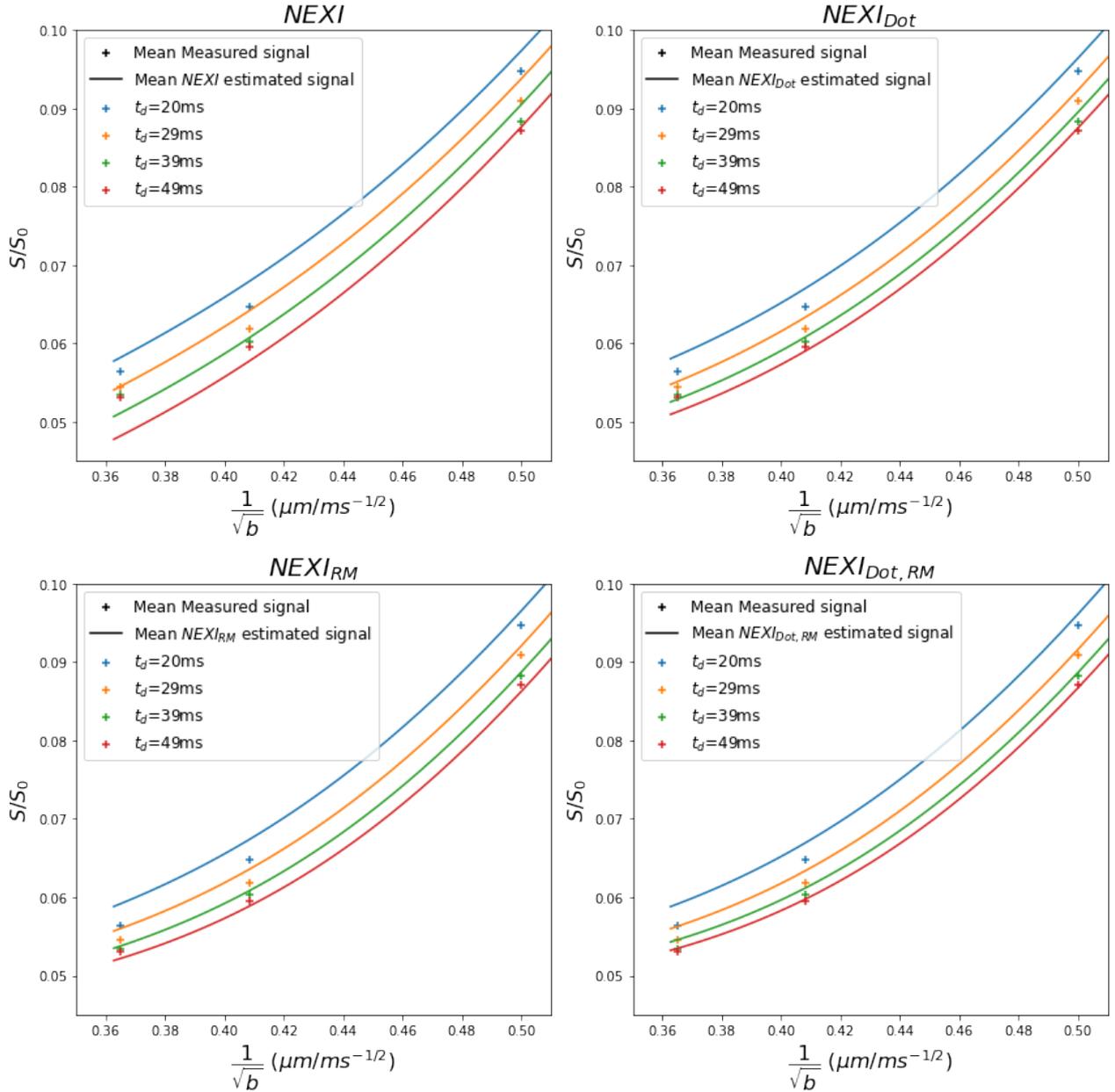

**Figure 3.** Mean estimated signal in the cortical ribbon by the four NEXI model variants at high b-value (b > 4 ms/μm²) compared to the mean measured signal, represented by plus signs. Each color represents a different diffusion time.

Furthermore, the dot fraction $f_{dot}$ estimated using $NEXI_{dot}$ was perfectly correlated with the Rician expectation value $\sqrt{\frac{\pi}{2}}\sigma$ in each ROI (**Figure 4**). The Kolmogorov-Smirnov (KS) test reveals that $f_{dot}$ and $\sigma$ distributions are similar ($p$ = 0.1967). This suggests that the dot compartment in $NEXI_{dot}$ is fitting the Rician floor with a systematic offset, casting doubt on an actual dot compartment being relevant for cortical GM in vivo, in agreement with (Tax et al., 2020) and that the $NEXI_{RM}$ model should therefore be preferred.

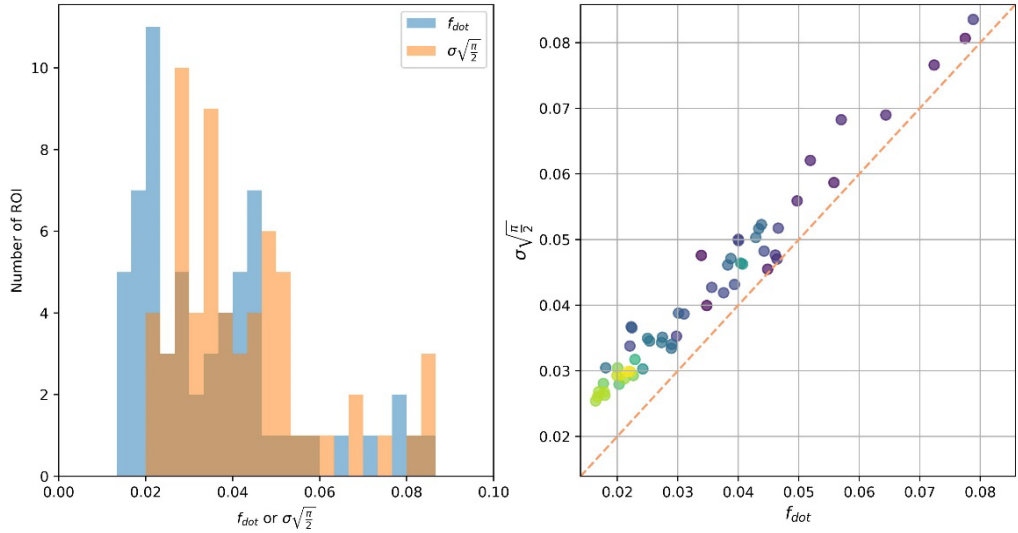

**Figure 3.** Agreement between $f_{dot}$ estimate of NEXI$_{dot}$ and the Rician floor expectation value, derived from the noise standard deviation (σ) obtained by denoising small b-values during preprocessing.

Based on this model variant comparison which favors the use of NEXI$_{RM}$ in vivo, we report NEXI gray matter microstructure estimates in the human brain (**Figure 5**). Using the NEXI$_{RM}$ implementation, quantitative maps show, as expected, $t_{ex}$ estimates in the range 20 – 50 ms in the cortex, and much longer in the white matter, where the diffusion time range does not allow a reliable estimation. The $D_e$ map reveals lower values in the cortex compared to sub-cortical white matter, which would be consistent with the high cellular abundance and random neurite orientations in GM slow down extra-cellular diffusion as compared to WM where diffusion at least along axons is less impeded. The $D_e$ contrast may also be consistent with the soma compartment being absorbed into the extra-cellular compartment in NEXI, thereby reducing its apparent diffusion in GM by the inclusion of restricted components. The neurite density fraction map reveals expected WM/GM contrast, with much higher fraction in WM; the cortical neurite fraction is estimated at ~40%. It should be noted that NEXI is not designed for WM, where the assumption of randomly oriented sticks and isotropic extra-neurite diffusivity is not expected to hold. This could have affected estimates in single-fiber WM population voxels vs crossing fiber WM areas, for example.

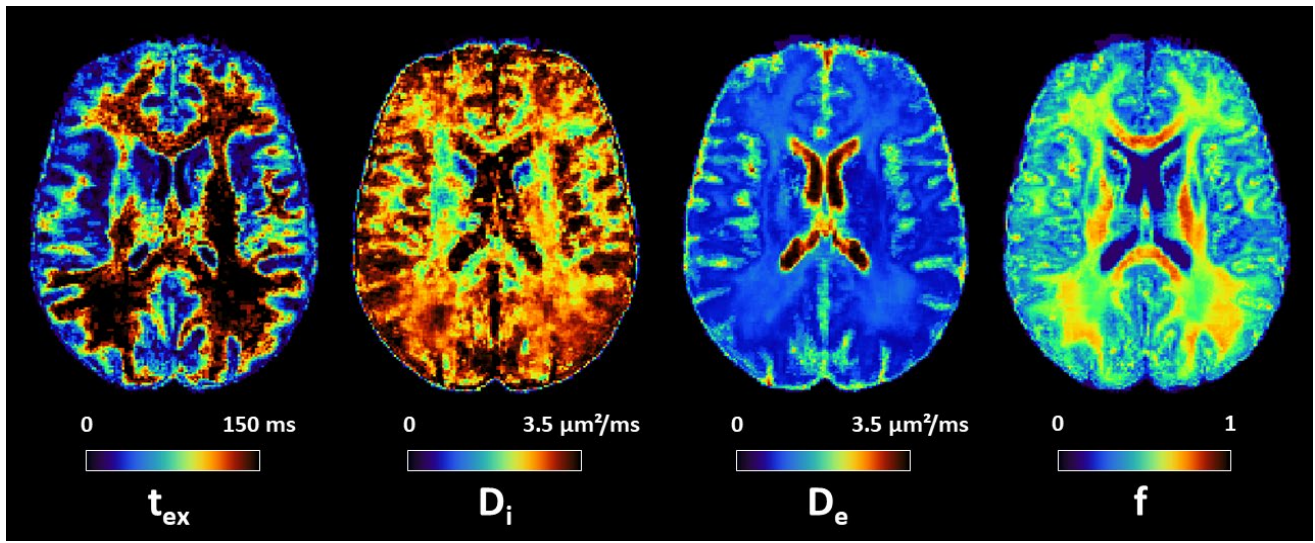

**Figure 5.** Axial slice of NEXI$_{RM}$ parametric maps, averaged across sessions and subjects (N=7). $t_{ex}$ and $D_e$ are consistent throughout the cortex, but $t_{ex}$ is presumably longer in the WM and cannot be reliably estimated using available diffusion times. $f$ displays the expected anatomical pattern in white vs gray matter. $D_i$ shows large variability across voxels, while hitting its upper bound frequently.

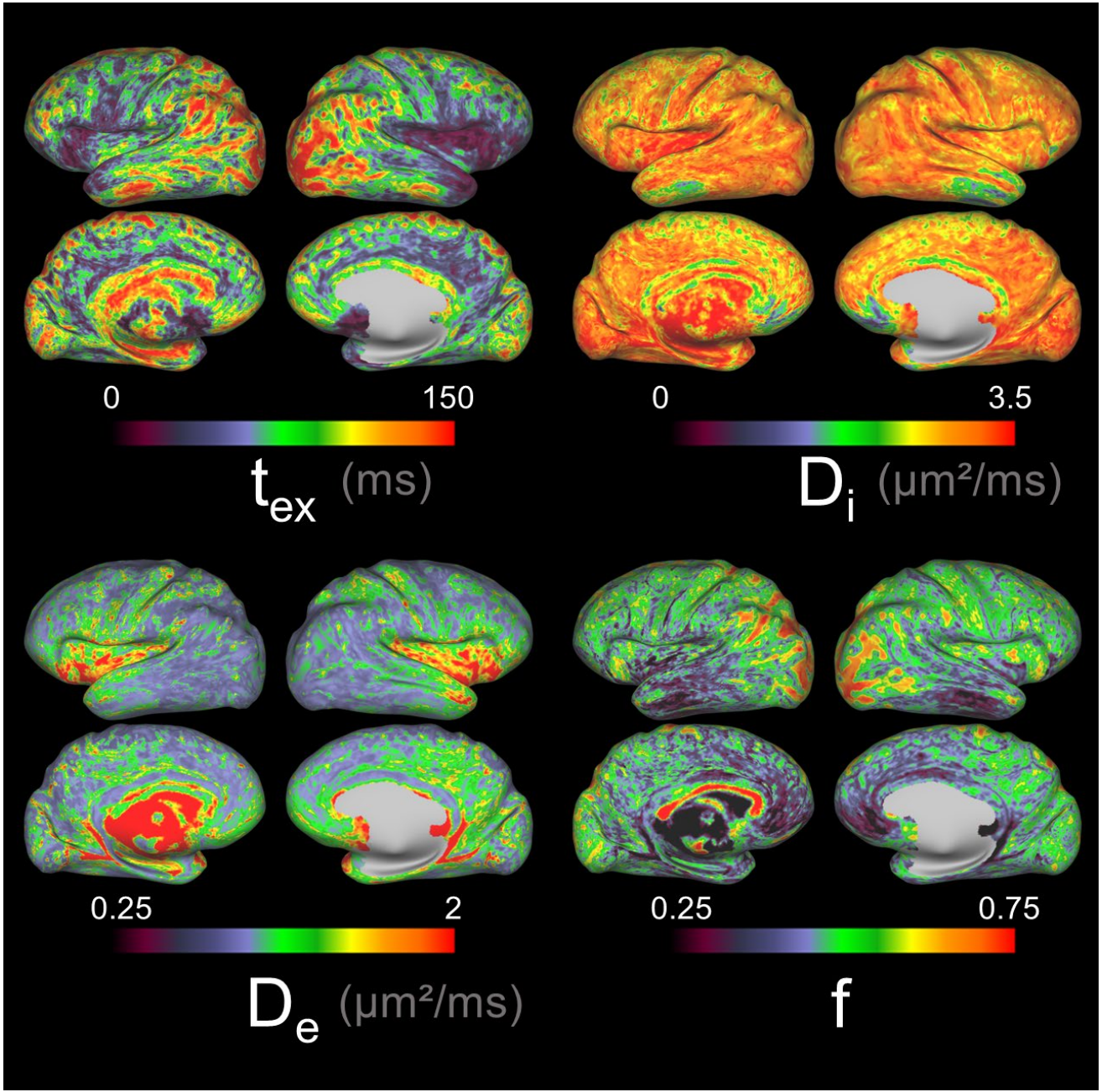

**Figure 6.** Projection onto cortical surface of NEXI$_{RM}$ maps averaged across subjects and sessions. We find some of the expected pattern of a larger fraction of neurites in the occipital lobe. This pattern can also be seen in the exchange time, which is also longer in the temporal lobe. A higher extracellular diffusivity is also observed in the somatosensory cortex.

These parametric maps, averaged within each DKT ROI, projected onto a study-average inflated cortical surface and averaged at the cortical thickness level voxel-wise after a multivariate template registration (Figure 6), reveal remarkable patterns across the healthy human brain. First, there is an expected level of symmetry between left and right hemispheres, although their estimates are completely independent, which suggests that spatial patterns are not casual.

**Figure 6** shows that the longest exchange time was found in the occipital lobe, in the posterior part of the parietal lobe and in the ventral parts of the temporal lobe, possibly indicating correlation with cortical myelination. $D_i$ estimates reach the upper bound in most of the regions of interest, limiting interpretation. However, a decrease in $D_i$ is observed in the rostral and ventral parts of the temporal lobe. $D_e$ revealed spatial patterns of faster extra-cellular diffusivity along the somatosensory cortex, as opposed to the occipital lobe and caudal part of the temporal lobe which have the slowest $D_e$. In the insula, $D_e$ is also considerably faster, however the level of partial volume effects might be higher, biasing the estimates upwards. As suggested above, $D_e$ is likely impacted by cellular density (extra-cellular tortuosity and high soma density) which reduces its estimate, or by fiber alignment that increases its estimate. Lastly, the neurite fraction $f$ follows a pattern of highest density in the occipital lobe and in the caudal part of the parietal lobe, comparable to $t_{ex}$ pattern possibly linked with myelination, but with moderate to lower densities in the ventral part of the temporal lobe. **Figure S3** in the supplementary material presents a comprehensive depiction of these results, showcasing the parametric medians per region of interest.

**Agreement with time-dependent diffusion and kurtosis**

Mean Diffusivity was almost independent of the diffusion time, with a weak yet measurable slope of $-7.5\times10^{-4}$ µm²/ms² ($p$=0.01) (**Figure 7A**). This weak diffusion time-dependence potentially calls into question the assumption of Gaussian compartments in our models and suggests structural disorder may introduce a confound (Lee et al., 2020). Mean Kurtosis decreased more markedly with time, which is consistent with previous studies (Jelescu et al., 2022; Lee et al., 2020). We find good agreement between $t_{ex}^{KM}$ obtained from MK(t) analysis and the one obtained from the NEXI$_{RM}$ fit. This agreement is expected as MK(t) in Eq (5) uses low b-value data that are less affected by Rician floor than the full NEXI model (Eqs. (1) and (3)).

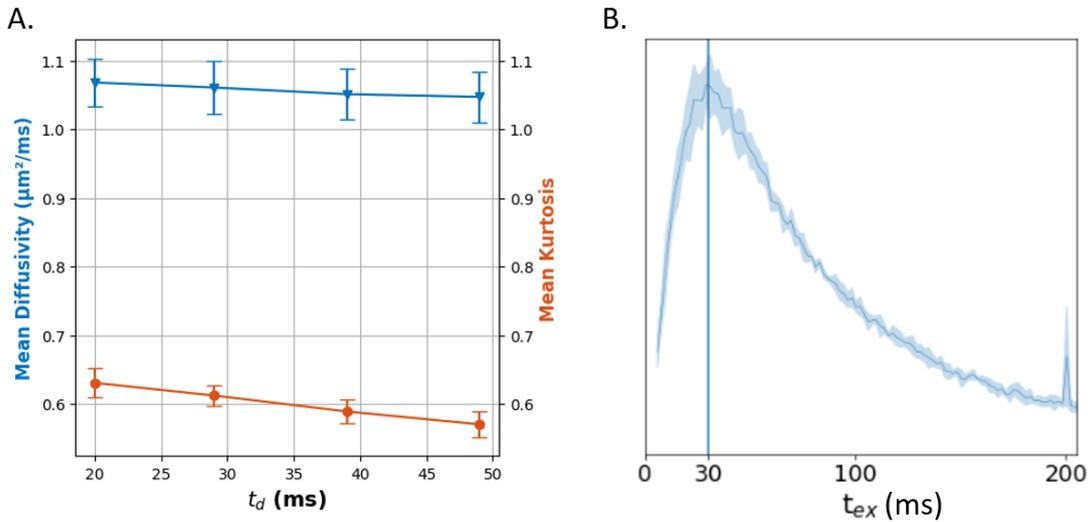

Figure 7. **A.** Time-dependent Mean Diffusivity and Mean Kurtosis in the cortex, averaged over voxels in the cortical ribbon and across the seven datasets (subjects and sessions). **B.** Distribution of $t_{ex}^{K(t)}$ values estimated voxelwise across the cortex, averaged across subjects (first session).

**Inter- vs intra-subject variability**

To assess intra-subject variability, we compared the first and second sessions of the three subjects who were scanned twice. To assess inter-subject variability, we compared the first session of the four subjects between them. Below, we compared NEXI$_{RM}$ results (**Figures 8 and 9**); for the other models, the plots are provided in **Supplementary Figure S4**.

The difference in median $t_{ex}$ over each ROI between different sessions is approximately 3.0 ms, while the difference in $t_{ex}$ across subjects is more than 2.5 times larger, at 7.70 ms (**Figure 8**). It is also noteworthy that the $t_{ex}$ do not display a broad range across the ROIs, with most values concentrated between 40 – 60 ms.

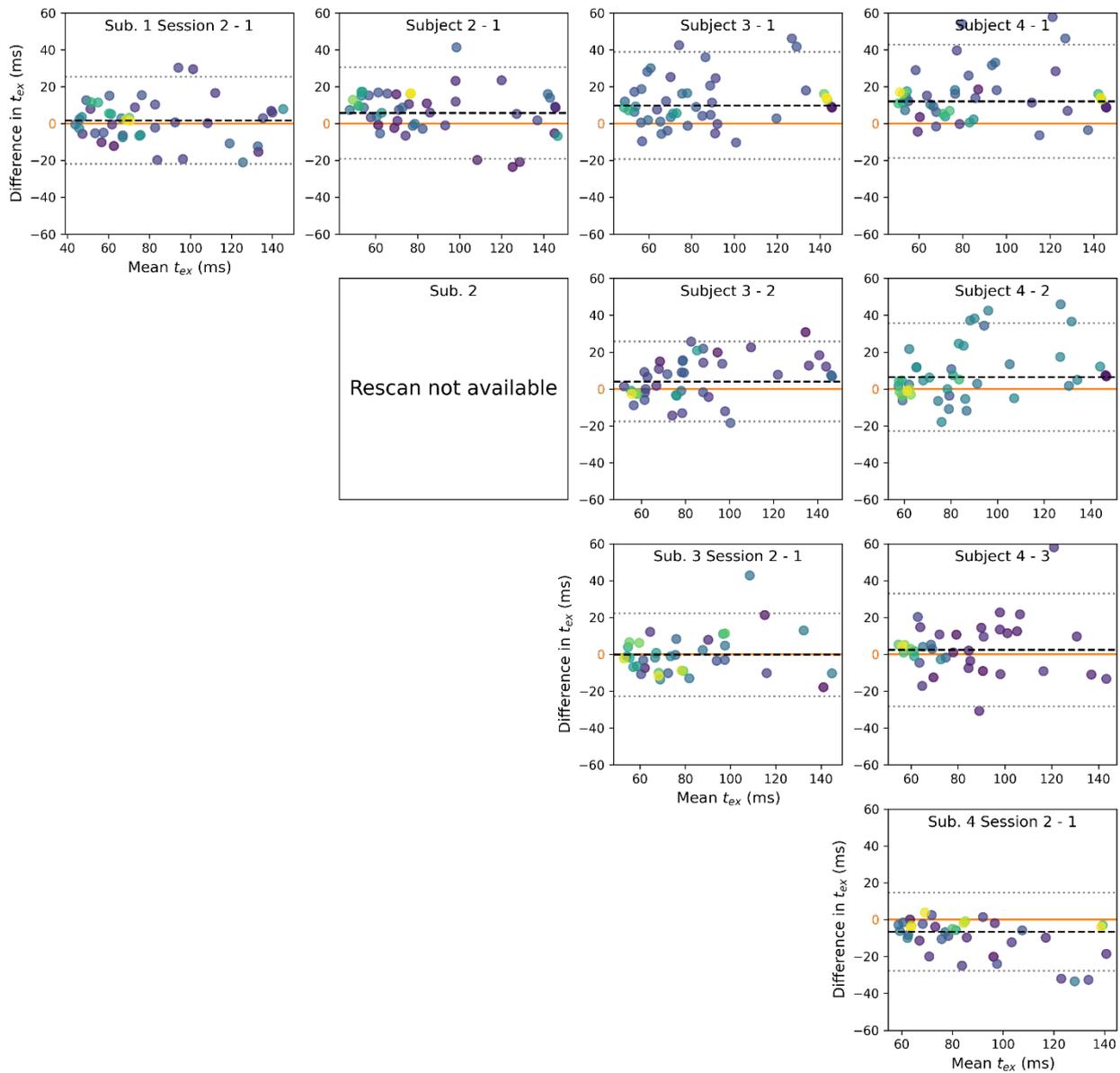

**Figure 8.** Bland-Altman plots on the $t_{ex}$ estimations from NEXI$_{RM}$ model. Each row and column refer to the same subject. On the diagonal, the two sessions of each subject are compared. In the upper triangle, the results of the first session of each subject are compared. The colors reflect the density of points in each range.

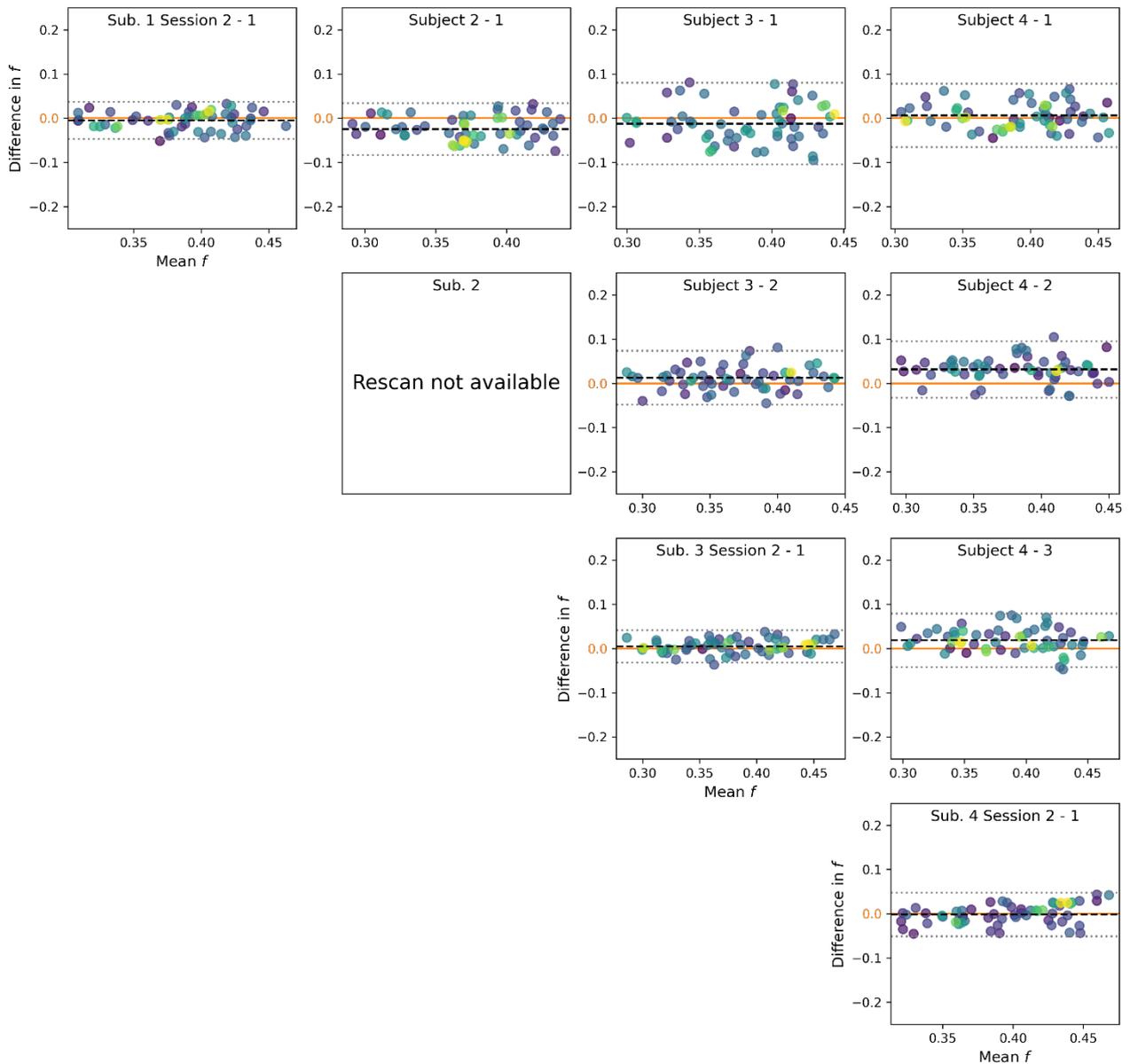

**Figure 9.** Bland-Altman plots on the $f$ estimations from NEXI$_{RM}$ model. Each row and column refer to the same subject. On the diagonal, the two sessions of each subject are compared. In the upper triangle, the results of the first session of each subject are compared. The colors reflect the density of points in each range.

In terms of neurite fraction $f$, the mean difference increases from 0.0040 for the inter-session comparisons to 0.01770 for the inter-subject comparisons (**Figure 9**), i.e. intra-subject variability is over four times larger than scan-rescan variability, a difference even more pronounced than for $t_{ex}$. The variance is also higher in the inter-subject vs intra-subject comparisons. Unlike the exchange time $t_{ex}$, neurite fraction values cover a broader range across DKT ROIs, showing brain regional specificity of this parameter.

This suggests that NEXI$_{RM}$ estimates are sufficiently reproducible to retain sensitivity to inter-subject differences.

For comparison, the Bland-Altman plots for $t_{ex}$ and $f$ of the other models can be found in Supplementary **Figure S4**. Additionally, the Bland-Altman plots for the two diffusivities of NEXI$_{RM}$ are available in Supplementary **Figure S5**.

## 4. Discussion

In this study, we compared different variants of the NEXI model in order to quantify microstructure features in the human cortex. We thus compared NEXI estimates, implemented as a two-compartment model with exchange as in (Jelescu et al., 2022), to those from its three-compartment variant NEXI$_{dot}$, also accounting for a dot compartment as proposed in (Olesen et al., 2022) for ex vivo data, as well as two new versions that correct for the Rician bias in the signal at high b-values: NEXI$_{RM}$ and NEXI$_{dot,RM}$. By examining these four model variants, the goal was to investigate the pertinence of a dot compartment to model human cortical gray matter, as the one shown in the cerebellum (Tax et al., 2020), and to study the effect of the Rician noise correction on these two models, given the lower SNR of clinical dMRI data as compared with preclinical data.

In the case where the ground truth is a two-compartment model with exchange and the standard deviation of the noise is known, the simulation results clearly show that the NEXI$_{RM}$ model is to be preferred against the other models and that the dot compartment is not able to substitute the Rician noise correction efficiently. Similarly, adding both a dot compartment and a Rician noise correction seems to disturb the model in the estimation of the main parameters, likely by the addition of an unnecessary free parameter ($f_{dot}$). The bias in NEXI estimates when the Rician floor is not accounted for is also very marked, although this bias is expected to be dependent on the SNR of the data. Simulations show that the performance of NEXI$_{RM}$ is equivalent to the performance of the other models in the case where the estimation of the noise level input into the Rician mean correction is overestimated by 50%.

While the RM correction is clearly beneficial, the performance of the three-compartment NEXI$_{dot}$ model on synthetic data generated using the two-compartment NEXI model is challenging to interpret. On the one hand, it is obvious that a non-zero dot compartment will be estimated, even when it is absent in the ground truth. On the other hand, although the existence of the dot compartment in healthy *in vivo* cerebrum tissue is not highly supported by histological evidence or previous experiments using spherical diffusion tensor encoding (Tax et al., 2020), we underline that our simulation results cannot reflect the performance of NEXI or NEXI$_{dot}$ in the case where a dot compartment would actually be present in the ground truth. However, NEXI$_{RM}$ and NEXI$_{dot}$ fits on experimental data show that NEXI$_{dot}$ essentially captures the Rician noise floor as a dot compartment, rather than the latter having a biological relevance as a compartment of its own. The slight but systematic lower level of the $f_{dot}$ estimate compared to the Rician floor $\sigma\sqrt{\frac{\pi}{2}}$ could be explained by the fact that the Rician correction is adaptive, mainly changing the signal magnitude at low SNR (high b-values) while the dot compartment acts by design as an offset to the signal across the entire b-value range. Thus, the $f_{dot}$ estimate is likely lower than the Rician floor as a compromise in fitting the signal well at both low and high b-values, in an MSE sense.

Looking strictly at the AICc of the four model fits on experimental data, one might be tempted to think that the NEXI$_{dot}$ model outperforms the other models. However, the AICc is also to be taken with caution because, voxelwise, $f_{dot}$ was a free parameter to fit, while σ was fixed to the value given by MP-PCA denoising during pre-processing. A potential error on σ could have reduced the performance of the NEXI$_{RM}$ fit, as also shown in the simulations of an overestimated σ, leading to a higher AICc. Therefore, the AICc should be compared between models with or without Rician mean correction. What these results then show is that when there is no Rician correction, the addition of a dot compartment better explains our data, but when the correction is added, it conversely becomes disadvantageous to include one.

Overall, our results on both synthetic and experimental data therefore indicate that the NEXI$_{RM}$ model, that is NEXI corrected for Rician noise, should be preferred for in vivo human cortex. It is noteworthy that the dot compartment may nonetheless be relevant as a biological compartment of its own in ex vivo data (Olesen et al., 2022). Furthermore, a soma compartment may be needed to better account for the signal decay at high b-values, although a model accounting for both soma and exchange (such as SANDIX) would likely require more datapoints and high SNR to yield reliable fit estimates, as discussed in the limitations paragraph.

The NEXI$_{RM}$ average estimate of $t_{ex}$ in the human cortical ribbon is 42 ms, vs 104 ms for NEXI. The former matches well with the average $t_{ex}^{K(t)}$ of 30 ms from the time-dependent kurtosis analysis, which is expected since $t_{ex}^{K(t)}$ is derived from data with b ≤ 2.5 ms/µm² which have higher SNR and are thus less impacted by the Rician floor. It is remarkable how different $t_{ex}$ estimates are across the four model variants, with the inclusion of a dot compartment systematically reducing $t_{ex}$. While the ground truth *in vivo* is not known, simulations support the experimental ordering in $t_{ex}$ estimates across models, with NEXI yielding the highest (and over-estimating $t_{ex}$ in simulations), followed by NEXI$_{RM}$ (with best accuracy in simulations), and finally NEXI$_{dot}$ and NEXI$_{dot,RM}$. The difference between the latter two is more pronounced in experimental data than in simulations, which could be attributed to partial volume effects or other tissue compartments not accounted for in the models, and that were absent in the simulations. Overall, these discrepancies reinforce the need to make informed decisions when selecting the model, as these decisions have a dramatic impact on ensuing exchange time estimates.

The estimation of other parameters is less variable across models. Extra-neurite diffusivity, $D_e$, is 0.9 – 1 µm²/ms, slightly higher than that reported in rats using NEXI (Jelescu et al., 2022). The neurite fraction, $f$, is ~0.3 – 0.4, also consistent with what has been reported in rats using NEXI (Jelescu et al., 2022). However, the obtained neurite volume fraction from histology, approximately 60% in the rat cortex (Braitenberg and Schüz, 1998; Chklovskii et al., 2002; Ikari and Hayashi, 1981), is much larger. This discrepancy may be due to a relaxation bias, if intra-neurite $T_2$ were shorter this would cause an underestimation of the compartment's volume fraction, or to faster exchange processes not captured by $t_{ex}$, that would also result in an underestimation of the restricted stick population.

However, for all models, the intra-neurite diffusivity measure, $D_i$, is unrealistically high, even above water diffusion coefficient at the body temperature of 3 µm²/ms, and often hits the upper bound implemented in the NLS algorithm. Intra-neurite or intra-axonal diffusivity is notoriously challenging to estimate, particularly in the presence of noise (Howard et al., 2022; Jelescu et al., 2016; Palombo et al., 2020). One possibility is that larger *b*-values combined with b-tensor encoding and/or $T_2$ relaxometry would be required to estimate $D_i$ in gray matter, as was the case in white matter (Dhital et al., 2019; Lampinen et al., 2020). Alternatively, working with real-valued data instead of magnitude data could help eliminate Rician bias and boost the SNR, thereby improving the $D_i$ estimates (Howard et al., 2022). It is also possible that a Partial Volume effect (PVE) takes place, where gray matter, white matter and CSF are captured in each voxel in varying proportions. This would make the model less suitable for our experimental data, pushing $D_i$ estimates towards unphysical values; the issue of PVE is discussed further below.

Based on the few recent works on gray matter exchange models, there seem to be dramatic differences in cortical gray matter microstructure features between *in vivo* and *ex vivo* tissue. Reported exchange times *ex vivo* are much shorter than in vivo, 3 – 14 ms irrespective of the inclusion or not of a dot compartment, neurite fractions are much higher 0.7 – 0.8, closer to their histological estimates (also based on ex vivo tissue) (Hertanu et al., 2023; Jelescu and Uhl, 2022; Olesen et al., 2022), contributions from structural disorder are more pronounced (Jelescu and Uhl, 2022), and $D_i$ is reduced within biologically plausible ranges (Hertanu et al.,

2023; Jelescu and Uhl, 2022). However, other groups have reported very short exchange times (3 – 10 ms) also in perfused viable rat pup spinal cord (Williamson et al., 2023, 2019), and even in human cortex in vivo (Lee et al., 2022), which would however translate into higher membrane permeability than ever reported for human neurons and astrocytes, as previously discussed (Boss et al., 2013; Jelescu et al., 2022). Similarly, previous works have put forward that structural disorder dominates over exchange in human cortex (Lee et al., 2020), considering detectable time-dependent diffusion in some ROIs. Here we also report weak yet detectable time-dependent diffusion when averaging across all voxels in the cortical ribbon and across subjects, with a significant negative slope. However, this contribution seems limited as compared to the exchange that drives a pronounced time-dependent kurtosis. Time-dependent diffusion in human cortex may also result from PVE with subcortical WM, as it was previously unambiguously reported in human WM (Fieremans et al., 2016), but not in the rat cortex in vivo where PVE with WM could be excluded (Jelescu et al., 2022).

For the first time, we also report NEXI parameter distributions across the surface of the human brain. While the maps of neurite fraction and exchange time do not fully align with the expected cortical myelin density mapping (Ali et al., 2022; Van Essen et al., 2018), brain regional differences are still in line with known variation in cell density and myelination across the cortex. Overall, longer exchange times, higher neurite densities and faster extracellular diffusivity (suggesting a more coherent alignment of neuronal processes) were found in motor, somatosensory and visual areas. It should also be underlined that biophysical models of water diffusion do not provide cell-type specific information, and astrocyte distributions across the cortex may also impact the NEXI maps in terms of "neurite density" (which rather mirrors cell process density) and exchange time (assuming astrocytes may be more permeable than neurons due to the presence of aquaporin-4 channels (Boss et al., 2013; Gleiser et al., 2016; Halnes et al., 2013)).

Finally, NEXI$_{RM}$ estimates display good scan-rescan repeatability while retaining sensitivity to inter-subject differences. These results are promising from the perspective of further clinical translation of NEXI, and its application to larger populations of healthy subjects and patients. The NEXI implementation on the Connectom scanner is a steppingstone between preclinical MRI systems and widespread clinical MRI systems. The advent of new human scanners featuring gradient amplitudes of 200 mT/m, such as the Cima.X (Siemens Healthineers) or the MAGNUS (GE Healthcare) (Foo et al., 2020), suggests that the next generation of MRI scanners will increasingly resemble the scanner used in this study, thereby expanding its scope. The potential of NEXI$_{RM}$ to estimate cortical microstructure features on a clinical scanner (Uhl et al., 2023) will be strengthened by the results of the present study, as it highlights the importance of correcting for Rician noise in the NEXI model to obtain accurate estimates of microstructure parameters in the human cortex. The Rician mean correction is expected to have even more influence on clinical data with lower SNR (due to the longer TE driven by weaker gradients up to 80 mT/m). The progress in hardware technology, exemplified by advancements like Connectom 2.0 (Huang et al., 2021), as well as the aforementioned new scanners, also holds significant promise for advancing the validation of reproducibility in this study at higher SNR and facilitating future clinical translation.

Our study has some limitations that should be noted. First, this study was a proof of principle, for which we sampled four participants. Future studies with larger sample sizes, possibly including patients, are warranted. Second, several trends suggest that the NEXI$_{RM}$ model, though more appropriate than the other three variants, may be incomplete to fully characterize cortical GM signal behavior in (q,t) space. The weak decay of D(t) may indicate that the model assumption of Gaussian compartments does not hold entirely; this should be further investigated on a larger cohort with a broader range of diffusion times. However, accounting for structural disorder explicitly in a biophysical model in combination with exchange is still work in progress for the community (Burcaw et al., 2015; Novikov et al., 2014). Furthermore, the soma compartment was neglected from the model, in light of the more pronounced effect of exchange over restriction signified by decreasing signal with increasing diffusion time (Jelescu et al., 2022; Olesen et al., 2022), but should represent a priority for future work. Indeed, quantifying soma at short diffusion times using SANDI has demonstrated value (Palombo et al., 2020) but is also challenging from the perspective of model degeneracy when combined with an exchange model as in SANDIX (Olesen et al., 2022). Recent approaches using different gradient waveforms have been proposed to separate the contributions of exchange (permeability) and restriction (soma) (Chakwizira et al., 2023) but led to much longer exchange time estimates than with NEXI, rather in line with previous literature using FEXI (Lampinen et al., 2017) which lacks specificity to biologically-relevant compartments. Structural disorder has also not been considered in this approach. Finally, residual effects of Rician noise may compromise the intra-neurite diffusivity estimate, which may benefit from working with real-valued vs magnitude data.

One of the advantages of the NEXI model is that it can be implemented on clinical scanners (Uhl et al., 2023) and thus enables studies in large cohorts of both healthy and patient populations. Future research will focus on its optimization on a clinical scanner with more moderate gradient set of 80 mT/m, although the availability of clinical scanners with 200 mT/m gradients can only ease the clinical translation of NEXI. Optimization avenues include accounting for the actual gradient pulse duration (as the narrow pulse approximation may not hold, as implemented in (Olesen et al., 2022)), trading magnitude data for real-valued data, trading NLS for a multi-layer perceptron fit and using explainable AI to optimize the clinical NEXI acquisition protocol within scanner hardware limits (Uhl et al., 2023). The main goal is to reduce both the acquisition time, and the estimation error on the two most challenging parameters, namely $D_i$ and $t_{ex}$. The development of a framework that enables joint estimation of soma and neurite permeability is also high priority.

## Conclusion

We reported the first comprehensive study of NEXI model parameter estimates in the human cortex in vivo. Our findings indicate that the addition of a dot compartment to the NEXI model is not necessary and that correcting the Rician floor in the fit is a more appropriate approach to account for its effects. The estimated exchange time, neurite fraction, and compartment diffusivities are consistent with previous studies conducted in the rat cortex in vivo, as well as with the exchange time estimate from time-dependent kurtosis. Notably, we observed that the exchange time is on the order of 30 – 40 ms, an intermediate value as compared to other similar studies but that signifies exchange cannot be neglected in the human GM at clinical diffusion times. These estimates displayed good scan-rescan repeatability, while preserving sensitivity to variations among subjects. However, the parameters $D_i$ and $t_{ex}$ were the most challenging to estimate, and future efforts will focus on possible improvements.

## Acknowledgments

QU, TP and IJ are supported by SNSF Eccellenza grant PCEFP2_194260. MP is supported by UKRI Future Leaders Fellowship MR/T020296/2. The data were acquired at the UK National Facility for In Vivo MR Imaging of Human Tissue Microstructure funded by the EPSRC (grant EP/M029778/1), and The Wolfson Foundation. The work is supported in part by a Wellcome Trust Investigator Award (096646/Z/11/Z) and Wellcome Trust Strategic Award (104943/Z/14/Z). For the purpose of open access, the author has applied a CC BY public copyright licence to any Author Accepted Manuscript version arising from this submission.

# References

Akaike, H. (Ed.), 1973. Information theory and an extension of the maximum likelihood principle. Akadémiai Kiadó, Budapest, Hungary.

Alexander, D.C., Dyrby, T.B., Nilsson, M., Zhang, H., 2019. Imaging brain microstructure with diffusion MRI: practicality and applications. NMR in Biomedicine 32, e3841. https://doi.org/10.1002/nbm.3841

Ali, T.S., Lv, J., Calamante, F., 2022. Gradual changes in microarchitectural properties of cortex and juxtacortical white matter: Observed by anatomical and diffusion MRI. Magnetic Resonance in Medicine 88, 2485–2503. https://doi.org/10.1002/mrm.29413

Altman, D.G., Bland, J.M., 1983. Measurement in Medicine: The Analysis of Method Comparison Studies. Journal of the Royal Statistical Society. Series D (The Statistician) 32, 307–317. https://doi.org/10.2307/2987937

Andersson, J.L.R., Sotiropoulos, S.N., 2016. An integrated approach to correction for off-resonance effects and subject movement in diffusion MR imaging. Neuroimage 125, 1063–1078. https://doi.org/10.1016/j.neuroimage.2015.10.019

Boss, D., Kühn, J., Jourdain, P., Depeursinge, C.D., Magistretti, P.J., M.d, P.P.M., 2013. Measurement of absolute cell volume, osmotic membrane water permeability, and refractive index of transmembrane water and solute flux by digital holographic microscopy. JBO 18, 036007. https://doi.org/10.1117/1.JBO.18.3.036007

Braitenberg, V., Schüz, A., 1998. Density of Dendrites, in: Braitenberg, V., Schüz, A. (Eds.), Cortex: Statistics and Geometry of Neuronal Connectivity. Springer, Berlin, Heidelberg, pp. 57–57. https://doi.org/10.1007/978-3-662-03733-1_11

Burcaw, L.M., Fieremans, E., Novikov, D.S., 2015. Mesoscopic structure of neuronal tracts from time-dependent diffusion. NeuroImage 114, 18–37. https://doi.org/10.1016/j.neuroimage.2015.03.061

Chakwizira, A., Zhu, A., Foo, T., Westin, C.-F., Szczepankiewicz, F., Nilsson, M., 2023. Diffusion MRI with free gradient waveforms on a high-performance gradient system: Probing restriction and exchange in the human brain. https://doi.org/10.48550/arXiv.2304.02764

Chklovskii, D.B., Schikorski, T., Stevens, C.F., 2002. Wiring optimization in cortical circuits. Neuron 34, 341–7.

Dhital, B., Reisert, M., Kellner, E., Kiselev, V.G., 2019. Intra-axonal diffusivity in brain white matter. NeuroImage 189, 543–550. https://doi.org/10.1016/j.neuroimage.2019.01.015

Fieremans, E., Burcaw, L.M., Lee, H.-H., Lemberskiy, G., Veraart, J., Novikov, D.S., 2016. In vivo observation and biophysical interpretation of time-dependent diffusion in human white matter. NeuroImage 129, 414–427. https://doi.org/10.1016/j.neuroimage.2016.01.018

Fieremans, E., Novikov, D.S., Jensen, J.H., Helpern, J.A., 2010. Monte Carlo study of a two-compartment exchange model of diffusion. NMR in biomedicine 23, 711–24. https://doi.org/10.1002/nbm.1577

Fieremans, Els, Novikov, D.S., Jensen, J.H., Helpern, J.A., 2010. Monte Carlo study of a two-compartment exchange model of diffusion. NMR in Biomedicine 23, 711–724.

Foo, T.K.F., Tan, E.T., Vermilyea, M.E., Hua, Y., Fiveland, E.W., Piel, J.E., Park, K., Ricci, J., Thompson, P.S., Graziani, D., Conte, G., Kagan, A., Bai, Y., Vasil, C., Tarasek, M., Yeo, D.T.B., Snell, F., Lee, D., Dean, A., DeMarco, J.K., Shih, R.Y., Hood, M.N., Chae, H., Ho, V.B., 2020. Highly efficient head-only magnetic field insert gradient coil for achieving simultaneous high gradient amplitude and slew rate at 3.0T (MAGNUS) for brain microstructure imaging. Magnetic Resonance in Medicine 83, 2356–2369. https://doi.org/10.1002/mrm.28087

Glasser, M.F., Coalson, T.S., Harms, M.P., Xu, J., Baum, G.L., Autio, J.A., Auerbach, E.J., Greve, D.N., Yacoub, E., Van Essen, D.C., Bock, N.A., Hayashi, T., 2022. Empirical transmit field bias correction of T1w/T2w myelin maps. NeuroImage 258, 119360. https://doi.org/10.1016/j.neuroimage.2022.119360

Gleiser, C., Wagner, A., Fallier-Becker, P., Wolburg, H., Hirt, B., Mack, A.F., 2016. Aquaporin-4 in Astroglial Cells in the CNS and Supporting Cells of Sensory Organs—A Comparative Perspective. International Journal of Molecular Sciences 17, 1411. https://doi.org/10.3390/ijms17091411

Halnes, G., Østby, I., Pettersen, K.H., Omholt, S.W., Einevoll, G.T., 2013. Electrodiffusive Model for Astrocytic and Neuronal Ion Concentration Dynamics. PLOS Computational Biology 9, e1003386. https://doi.org/10.1371/journal.pcbi.1003386

Henriques, R.N., Jespersen, S.N., Shemesh, N., 2019. Microscopic anisotropy misestimation in spherical-mean single diffusion encoding MRI. Magnetic resonance in medicine 81, 3245–3261.

Henschel, L., Conjeti, S., Estrada, S., Diers, K., Fischl, B., Reuter, M., 2020. FastSurfer - A fast and accurate deep learning based neuroimaging pipeline. NeuroImage 219, 117012. https://doi.org/10.1016/j.neuroimage.2020.117012

Hertanu, A., Uhl, Q., Pavan, T., Lamy, C.M., Jelescu, I.O., 2023. Quantifying features of human gray matter microstructure postmortem using Neurite Exchange Imaging (NEXI) at ultra-high field. Presented at the ISMRM Annual Meeting 2023, Toronto.

Howard, A.F., Cottaar, M., Drakesmith, M., Fan, Q., Huang, S.Y., Jones, D.K., Lange, F.J., Mollink, J., Rudrapatna, S.U., Tian, Q., Miller, K.L., Jbabdi, S., 2022. Estimating axial diffusivity in the NODDI model. NeuroImage 262, 119535. https://doi.org/10.1016/j.neuroimage.2022.119535

Huang, S.Y., Witzel, T., Keil, B., Scholz, A., Davids, M., Dietz, P., Rummert, E., Ramb, R., Kirsch, J.E., Yendiki, A., Fan, Q., Tian, Q., Ramos-Llordén, G., Lee, H.-H., Nummenmaa, A., Bilgic, B., Setsompop, K., Wang, F., Avram, A.V., Komlosh, M., Benjamini, D., Magdoom, K.N., Pathak, S., Schneider, W., Novikov, D.S., Fieremans, E., Tounekti, S., Mekkaoui, C., Augustinack, J., Berger, D., Shapson-Coe, A., Lichtman, J., Basser, P.J., Wald, L.L., Rosen, B.R., 2021. Connectome 2.0: Developing the next-generation ultra-high gradient strength human MRI scanner for bridging studies of the micro-, meso- and macro-connectome. Neuroimage 243, 118530. https://doi.org/10.1016/j.neuroimage.2021.118530

Ikari, K., Hayashi, M., 1981. Aging in the Neuropil of Cerebral Cortex– A Quantitative Ultrastructural Study. Psychiatry and Clinical Neurosciences 35, 477–486. https://doi.org/10.1111/j.1440-1819.1981.tb00245.x

Illán-Gala, I., Montal, V., Borrego-Écija, S., Mandelli, M.L., Falgàs, N., Welch, A.E., Pegueroles, J., Santos-Santos, M., Bejanin, A., Alcolea, D., Dols-Icardo, O., Belbin, O., Sánchez-Saudinós, M.B., Bargalló, N., González-Ortiz, S., Lladó, A., Blesa, R., Dickerson, B.C., Rosen, H.J., Miller, B.L., Lleó, A., Gorno-Tempini, M.L., Sánchez-Valle, R., Fortea, J., 2022. Cortical microstructure in primary progressive aphasia: a multicenter study. Alzheimer's Research & Therapy 14, 27. https://doi.org/10.1186/s13195-022-00974-0

Jelescu, I.O., de Skowronski, A., Geffroy, F., Palombo, M., Novikov, D.S., 2022. Neurite Exchange Imaging (NEXI): A minimal model of diffusion in gray matter with inter-compartment water exchange. NeuroImage 256, 119277. https://doi.org/10.1016/j.neuroimage.2022.119277

Jelescu, I.O., Palombo, M., Bagnato, F., Schilling, K.G., 2020. Challenges for biophysical modeling of microstructure. Journal of Neuroscience Methods 344, 108861. https://doi.org/10.1016/j.jneumeth.2020.108861

Jelescu, I.O., Uhl, Q., 2022. Ex vivo gray matter is complex: exchange & disorder exponents. Presented at the ISMRM.

Jelescu, I.O., Veraart, J., Fieremans, E., Novikov, D.S., 2016. Degeneracy in model parameter estimation for multi-compartmental diffusion in neuronal tissue: Degeneracy in Model Parameter Estimation of Diffusion in Neural Tissue. NMR Biomed. 29, 33–47. https://doi.org/10.1002/nbm.3450

Jensen, J.H., Helpern, J.A., 2010. MRI quantification of non-Gaussian water diffusion by kurtosis analysis. NMR in Biomedicine 23, 698–710.

Jensen, J.H., Helpern, J.A., Ramani, A., Lu, H., Kaczynski, K., 2005. Diffusional kurtosis imaging: The quantification of non-gaussian water diffusion by means of magnetic resonance imaging. Magn. Reson. Med. 53, 1432–1440. https://doi.org/10.1002/mrm.20508

Jones, D.K., Alexander, D.C., Bowtell, R., Cercignani, M., Dell'Acqua, F., McHugh, D.J., Miller, K.L., Palombo, M., Parker, G.J.M., Rudrapatna, U.S., Tax, C.M.W., 2018. Microstructural imaging of the human brain with a 'super-scanner': 10 key advantages of ultra-strong gradients for diffusion MRI. NeuroImage, Microstructural Imaging 182, 8–38. https://doi.org/10.1016/j.neuroimage.2018.05.047

Kärger, J., 1985. NMR self-diffusion studies in heterogeneous systems. Advances in Colloid and Interface Science 23, 129–148. https://doi.org/10.1016/0001-8686(85)80018-X

Kellner, E., Dhital, B., Kiselev, V.G., Reisert, M., 2016. Gibbs-ringing artifact removal based on local subvoxel-shifts. Magn Reson Med 76, 1574–1581. https://doi.org/10.1002/mrm.26054

Klein, A., Tourville, J., 2012. 101 Labeled Brain Images and a Consistent Human Cortical Labeling Protocol. Frontiers in Neuroscience 6.

Kunz, N., da Silva, A.R., Jelescu, I.O., 2018. Intra- and extra-axonal axial diffusivities in the white matter: Which one is faster? NeuroImage 181, 314–322. https://doi.org/10.1016/j.neuroimage.2018.07.020

Lampinen, B., Szczepankiewicz, F., Mårtensson, J., van Westen, D., Hansson, O., Westin, C.-F., Nilsson, M., 2020. Towards unconstrained compartment modeling in white matter using diffusion-relaxation MRI with tensor-valued diffusion encoding. Magnetic Resonance in Medicine 84, 1605–1623. https://doi.org/10.1002/mrm.28216

Lampinen, B., Szczepankiewicz, F., van Westen, D., Englund, E., C Sundgren, P., Lätt, J., Ståhlberg, F., Nilsson, M., 2017. Optimal experimental design for filter exchange imaging: Apparent exchange rate measurements in the healthy brain and in intracranial tumors. Magnetic Resonance in Medicine 77, 1104–1114. https://doi.org/10.1002/mrm.26195

Lee, H.H., Olesen, J.L., Tian, Q., Llorden, G.R., Jespersen, S.N., Huang, S.Y., 2022. Revealing diffusion time-dependence and exchange effect in the in vivo human brain gray matter by using high gradient diffusion. Presented at the ISMRM Annual Meeting 2022, London.

Lee, H.-H., Papaioannou, A., Novikov, D.S., Fieremans, E., 2020. In vivo observation and biophysical interpretation of time-dependent diffusion in human cortical gray matter. Neuroimage 222, 117054. https://doi.org/10.1016/j.neuroimage.2020.117054

Marcus, D., Harwell, J., Olsen, T., Hodge, M., Glasser, M., Prior, F., Jenkinson, M., Laumann, T., Curtiss, S., Van Essen, D., 2011. Informatics and Data Mining Tools and Strategies for the Human Connectome Project. Frontiers in Neuroinformatics 5.

McKinnon, E.T., Jensen, J.H., Glenn, G.R., Helpern, J.A., 2017. Dependence on b-value of the direction-averaged diffusion-weighted imaging signal in brain. Magnetic resonance imaging 36, 121–127.

Novikov, D.S., Fieremans, E., Jespersen, S.N., Kiselev, V.G., 2019. Quantifying brain microstructure with diffusion MRI: Theory and parameter estimation. NMR in Biomedicine 32, e3998.

Novikov, D.S., Jensen, J.H., Helpern, J.A., Fieremans, E., 2014. Revealing mesoscopic structural universality with diffusion. Proceedings of the National Academy of Sciences 111, 5088–5093. https://doi.org/10.1073/pnas.1316944111

Novikov, D.S., Kiselev, V.G., Jespersen, S.N., 2018. On modeling. Magnetic Resonance in Medicine 79, 3172–3193. https://doi.org/10.1002/mrm.27101

Nürnberger, L., Gracien, R.-M., Hok, P., Hof, S.-M., Rüb, U., Steinmetz, H., Hilker, R., Klein, J.C., Deichmann, R., Baudrexel, S., 2017. Longitudinal changes of cortical microstructure in Parkinson's disease assessed with T1 relaxometry. NeuroImage: Clinical 13, 405–414. https://doi.org/10.1016/j.nicl.2016.12.025

Olesen, J.L., Østergaard, L., Shemesh, N., Jespersen, S.N., 2022. Diffusion time dependence, power-law scaling, and exchange in gray matter. NeuroImage 251, 118976. https://doi.org/10.1016/j.neuroimage.2022.118976

Palombo, M., Ianus, A., Guerreri, M., Nunes, D., Alexander, D.C., Shemesh, N., Zhang, H., 2020. SANDI: A compartment-based model for non-invasive apparent soma and neurite imaging by diffusion MRI. NeuroImage 215, 116835. https://doi.org/10.1016/j.neuroimage.2020.116835

Palombo, M., Shemesh, N., Ianus, A., Alexander, D., Zhang, H., 2018. Abundance of cell bodies can explain the stick model's failure in grey matter at high bvalue [WWW Document]. In: Miller, KL and Port, JD, (eds.) Proceedings of the Joint Annual Meeting ISMRM-ESMRMB 2018. ISMRM (International Society for Magnetic Resonance in Medicine): Concord, CA, USA. (2018). URL https://www.ismrm.org/18m/ (accessed 2.11.22).

Setsompop, K., Kimmlingen, R., Eberlein, E., Witzel, T., Cohen-Adad, J., McNab, J.A., Keil, B., Tisdall, M.D., Hoecht, P., Dietz, P., Cauley, S.F., Tountcheva, V., Matschl, V., Lenz, V.H., Heberlein, K., Potthast, A., Thein, H., Van Horn, J., Toga, A., Schmitt, F., Lehne, D., Rosen, B.R., Wedeen, V., Wald, L.L., 2013. Pushing the limits of in vivo diffusion MRI for the Human Connectome Project. NeuroImage, Mapping the Connectome 80, 220–233. https://doi.org/10.1016/j.neuroimage.2013.05.078

Spotorno, N., Strandberg, O., Vis, G., Stomrud, E., Nilsson, M., Hansson, O., 2022. Measures of cortical microstructure are linked to amyloid pathology in Alzheimer's disease. Brain awac343. https://doi.org/10.1093/brain/awac343

Stanisz, G.J., Szafer, A., Wright, G.A., Henkelman, R.M., 1997. An analytical model of restricted diffusion in bovine optic nerve. Magnetic resonance in medicine : official journal of the Society of Magnetic Resonance in Medicine / Society of Magnetic Resonance in Medicine 37, 103–11.

Tax, C.M.W., Szczepankiewicz, F., Nilsson, M., Jones, D.K., 2020. The dot-compartment revealed? Diffusion MRI with ultra-strong gradients and spherical tensor encoding in the living human brain. NeuroImage 210, 116534. https://doi.org/10.1016/j.neuroimage.2020.116534

Uhl, Q., Pavan, T., Feiweier, T., Canales-Rodríguez, E.J., Jelescu, I.O., 2023. Optimizing the NEXI acquisition protocol for quantifying human gray matter microstructure on a clinical MRI scanner using Explainable AI. Presented at the ISMRM 2023, Toronto.

Van Essen, D.C., Donahue, C.J., Glasser, M.F., 2018. Development and Evolution of Cerebral and Cerebellar Cortex. Brain Behavior and Evolution 91, 158–169. https://doi.org/10.1159/000489943

Veraart, J., Fieremans, E., Novikov, D.S., 2016a. Universal power-law scaling of water diffusion in human brain defines what we see with MRI. https://doi.org/10.48550/arXiv.1609.09145

Veraart, J., Fieremans, E., Rudrapatna, U., Jones, D.K., Novikov, D.S., 2018. Biophysical modeling of the gray matter: does the "stick" model hold? Presented at the ISMRM, Paris, France, p. 1094.

Veraart, J., Novikov, D.S., Christiaens, D., Ades-aron, B., Sijbers, J., Fieremans, E., 2016b. Denoising of diffusion MRI using random matrix theory. Neuroimage 142, 394–406. https://doi.org/10.1016/j.neuroimage.2016.08.016

Veraart, J., Sijbers, J., Sunaert, S., Leemans, A., Jeurissen, B., 2013. Weighted linear least squares estimation of diffusion MRI parameters: Strengths, limitations, and pitfalls. NeuroImage 81, 335–346. https://doi.org/10.1016/j.neuroimage.2013.05.028

Virtanen, P., Gommers, R., Oliphant, T.E., Haberland, M., Reddy, T., Cournapeau, D., Burovski, E., Peterson, P., Weckesser, W., Bright, J., van der Walt, S.J., Brett, M., Wilson, J., Millman, K.J., Mayorov, N., Nelson, A.R.J., Jones, E., Kern, R., Larson, E., Carey, C.J., Polat, İ., Feng, Y., Moore, E.W., VanderPlas, J., Laxalde, D., Perktold, J., Cimrman, R., Henriksen, I., Quintero, E.A., Harris, C.R., Archibald, A.M., Ribeiro, A.H., Pedregosa, F., van Mulbregt, P., 2020. SciPy 1.0: fundamental algorithms for scientific computing in Python. Nat Methods 17, 261–272. https://doi.org/10.1038/s41592-019-0686-2

Voldsbekk, I., Bjørnerud, A., Groote, I., Zak, N., Roelfs, D., Maximov, I.I., Geier, O., Due-Tønnessen, P., Bøen, E., Kuiper, Y.S., Løkken, L.-L., Strømstad, M., Blakstvedt, T.Y., Bjorvatn, B., Malt, U.F., Westlye, L.T., Elvsåshagen, T., Grydeland, H., 2022. Evidence for widespread alterations in cortical microstructure after 32 h of sleep deprivation. Transl Psychiatry 12, 1–8. https://doi.org/10.1038/s41398-022-01909-x

Williamson, N.H., Ravin, R., Benjamini, D., Merkle, H., Falgairolle, M., O'Donovan, M.J., Blivis, D., Ide, D., Cai, T.X., Ghorashi, N.S., Bai, R., Basser, P.J., 2019. Magnetic resonance measurements of cellular and sub-cellular membrane structures in live and fixed neural tissue. eLife 8, e51101. https://doi.org/10.7554/eLife.51101

Williamson, N.H., Ravin, R., Cai, T.X., Falgairolle, M., O'Donovan, M.J., Basser, P.J., 2023. Water exchange rates measure active transport and homeostasis in neural tissue. PNAS Nexus 2, pgad056. https://doi.org/10.1093/pnasnexus/pgad056

# Quantifying human gray matter microstructure using NEXI and 300 mT/m gradients

*Quentin Uhl, Tommaso Pavan, Malwina Molendowska, Derek K. Jones, Marco Palombo, Ileana Jelescu*

## Supplementary Material

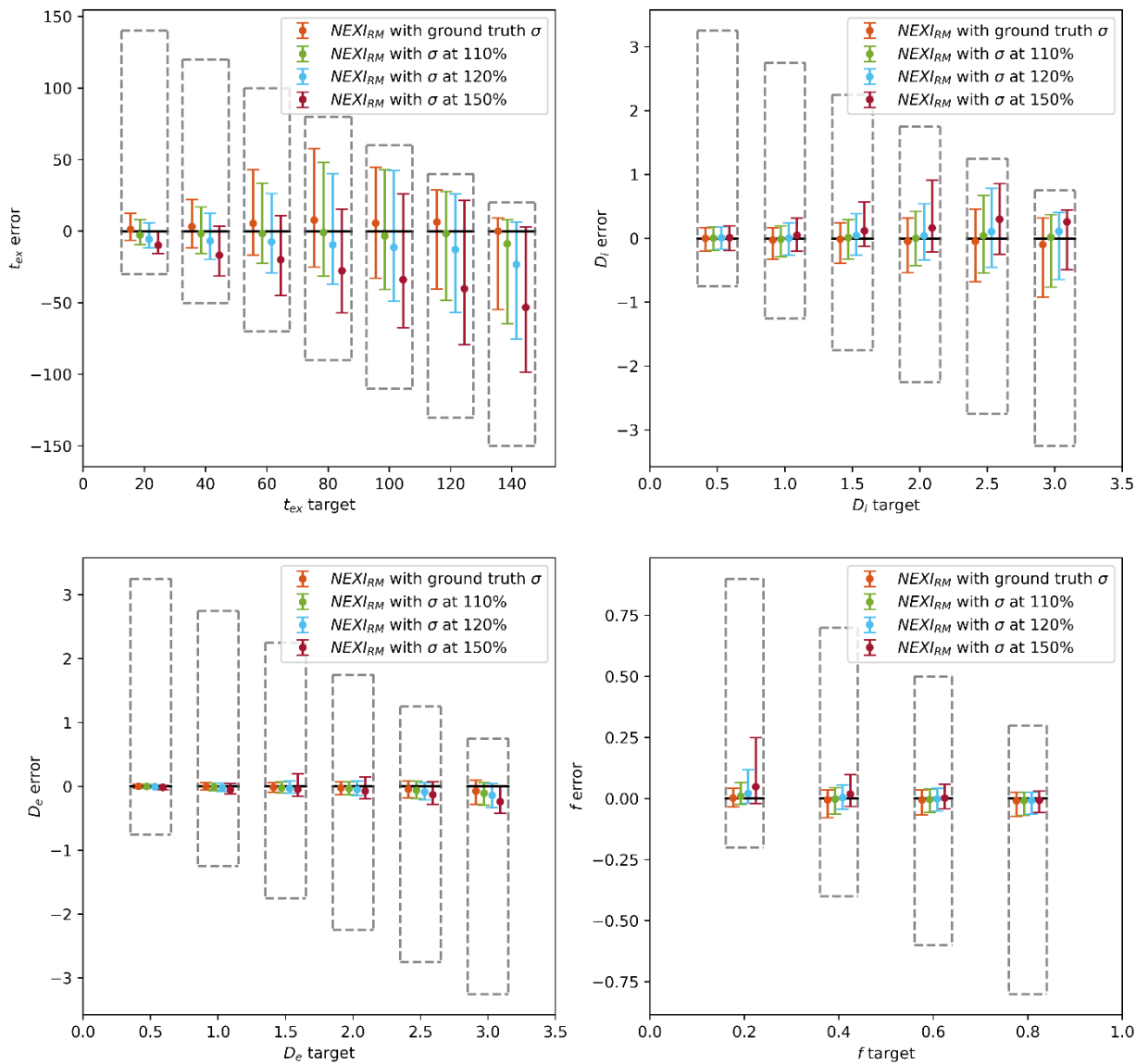

**Figure S1.** Boxplots (median and interquartile range) of parameter estimates by NEXI$_{RM}$ using the true and overestimated σ, on a synthetic dataset with random Rician noise (true σ=0.03). The error is defined as the difference between the estimation and the target value. The upper and lower limits of the grey dashed box represent the maximum and minimum error of the estimator, in the given bin, due to lower and upper bounds in the NLS algorithm.

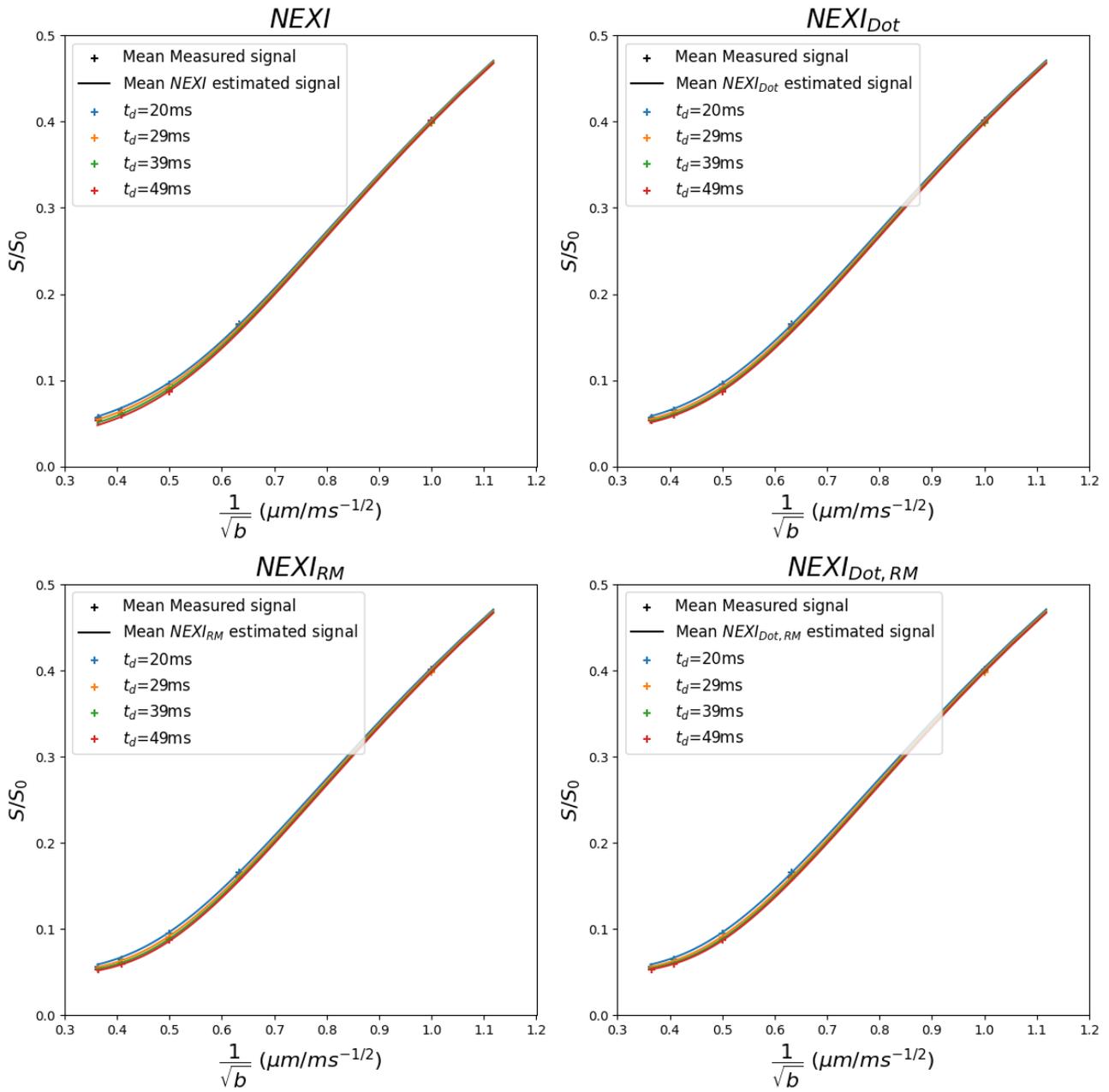

**Figure S2**. Mean estimated signal in the cortical ribbon by the four NEXI model variants at both high and low b-value compared to the mean measured signal, represented by plus signs. Each color represents a different diffusion time. The low b-value (especially at b=2.5ms/µm²) mean measured signal caused the estimated mean curve to shift upwards at high b-value.

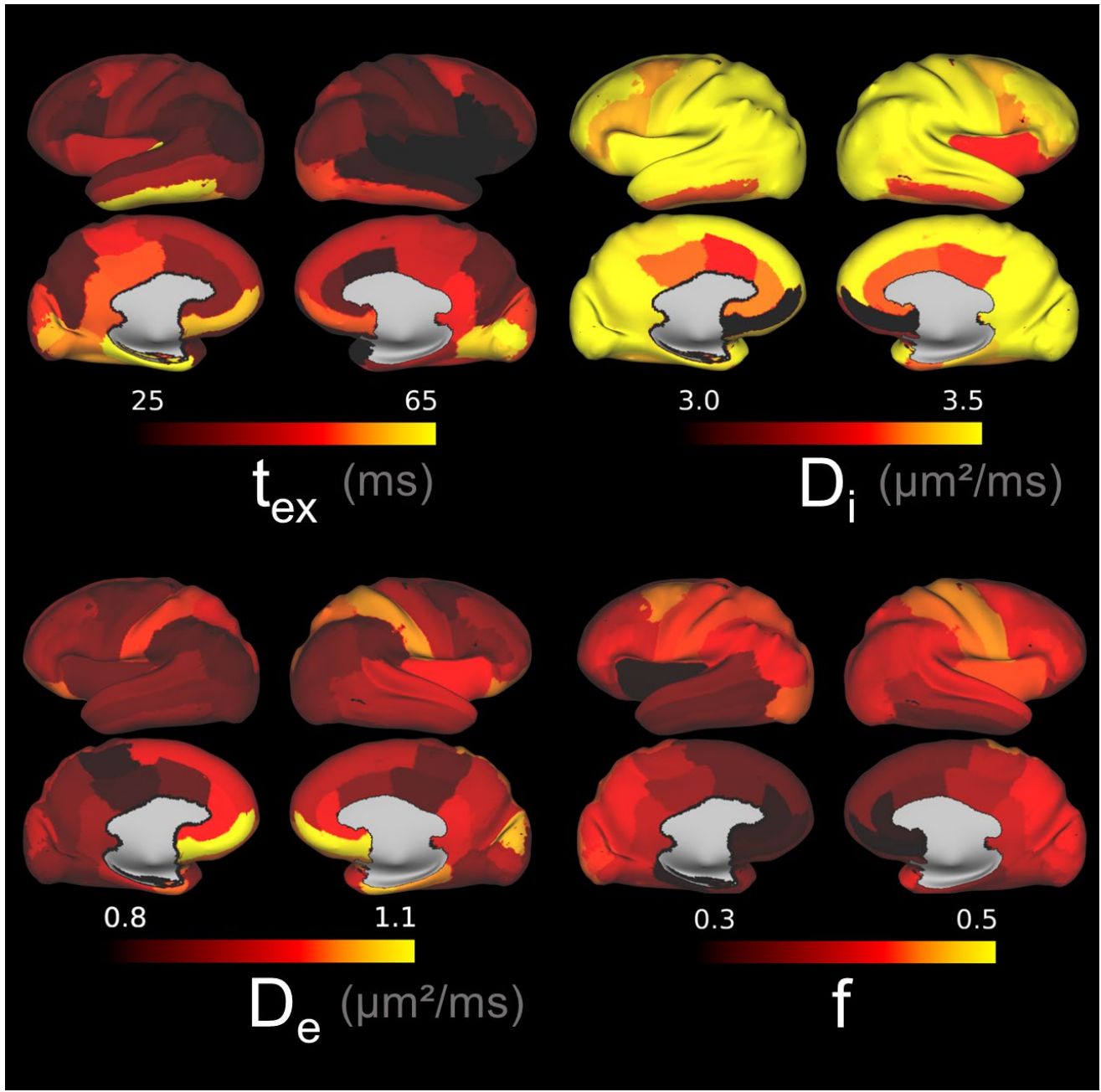

**Figure S3**. Average NEXI_RM estimates for each DKT ROI. The averages were computed from the volumetric maps and projected to the study-average inflated cortical surface. We find patterns of a larger fraction of neurites in the motor cortex and premotor area. The exchange time is longer in the temporal and occipital lobes, and in the premotor area. A faster extracellular diffusivity is observed in the somatosensory cortex.

**Figure S4.** Bland-Altman plots on the $t_{ex}$ and f estimations from NEXI (A-B), NEXI$_{dot}$ (C-D) and NEXI$_{dot,RM}$ (E-F) models. Each row and column refer to the same subject. On the diagonal, the two sessions of each subject are compared. In the upper triangle, the results of the first session of each subject are compared.

**A.**

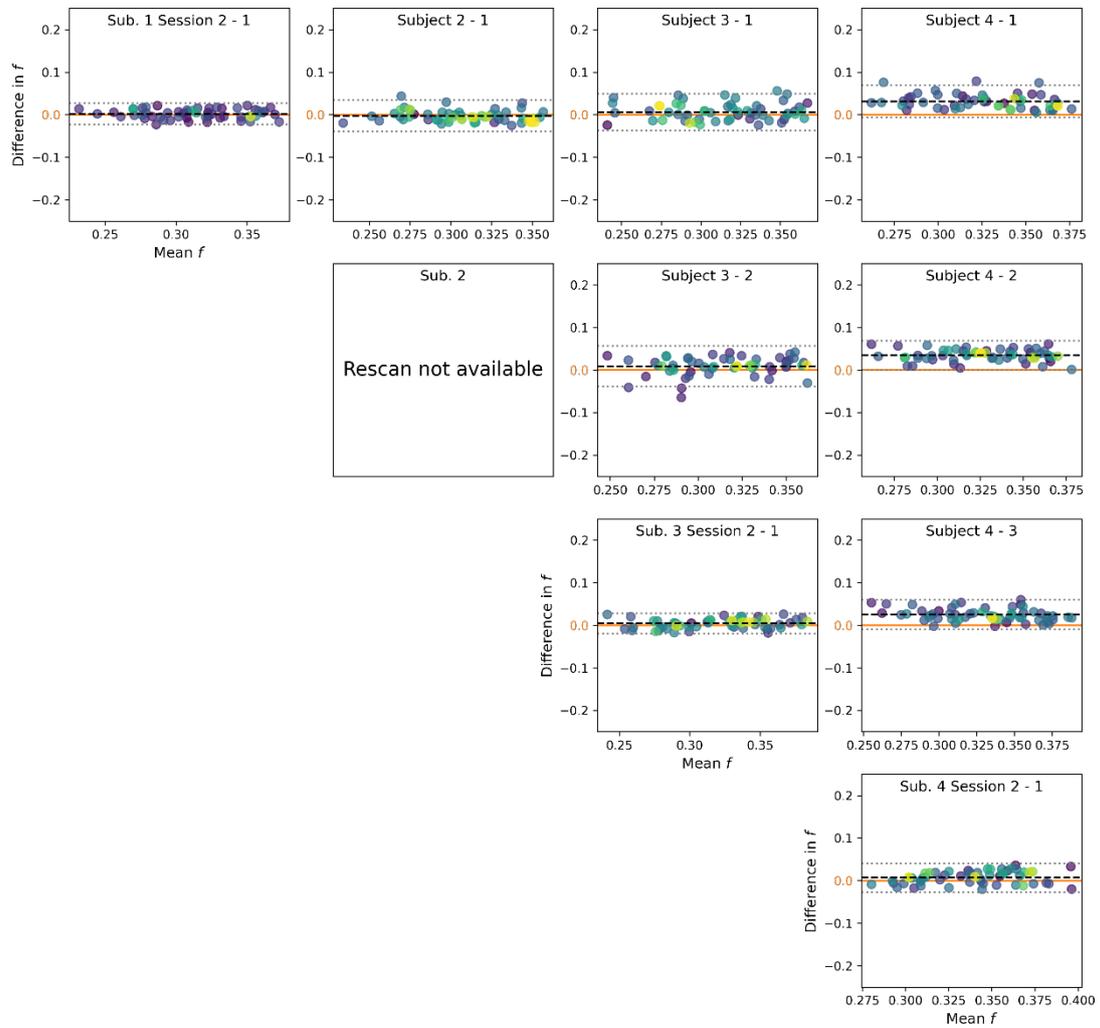

**B.**

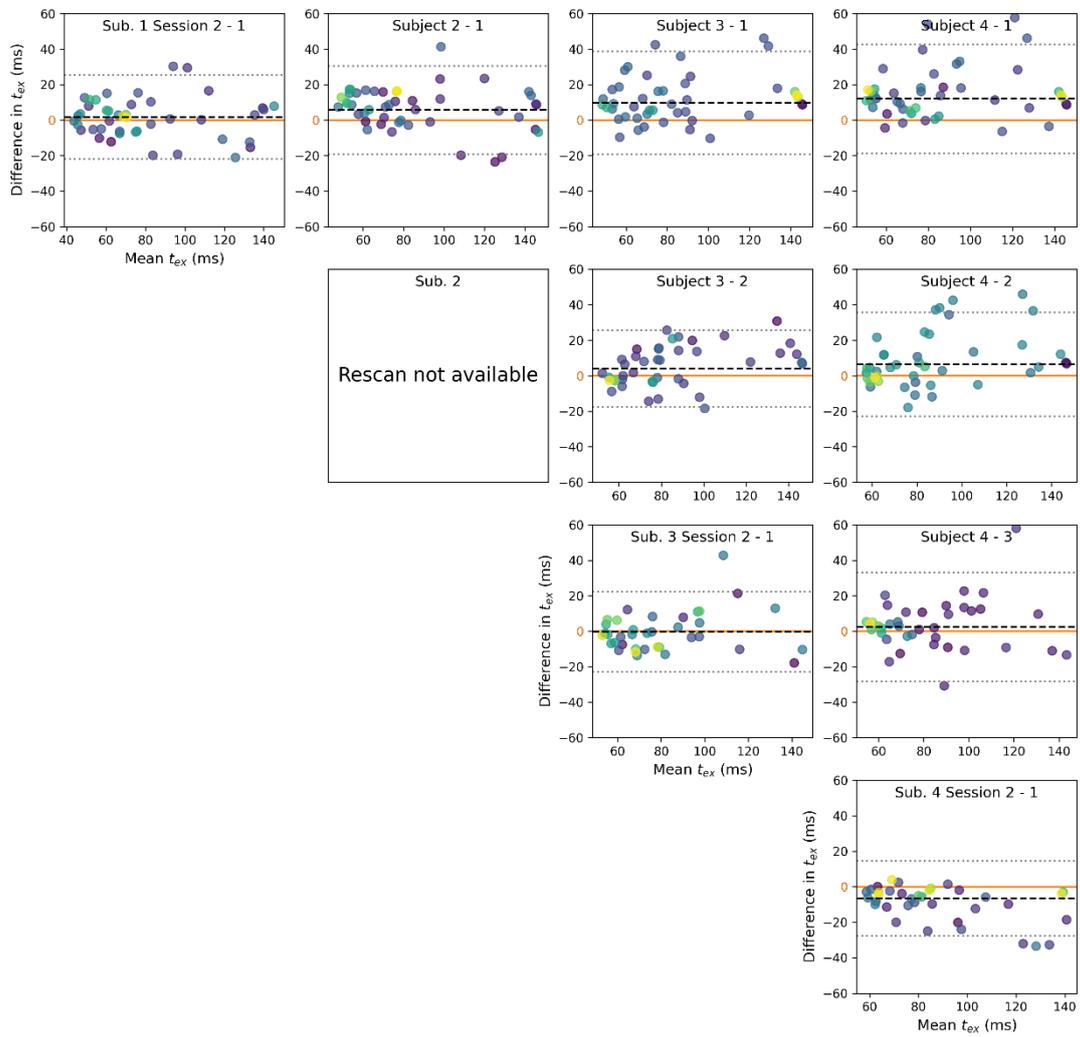

Bland-Altman plot of agreement between $t_{ex}$ median ROI estimations using *NEXI*

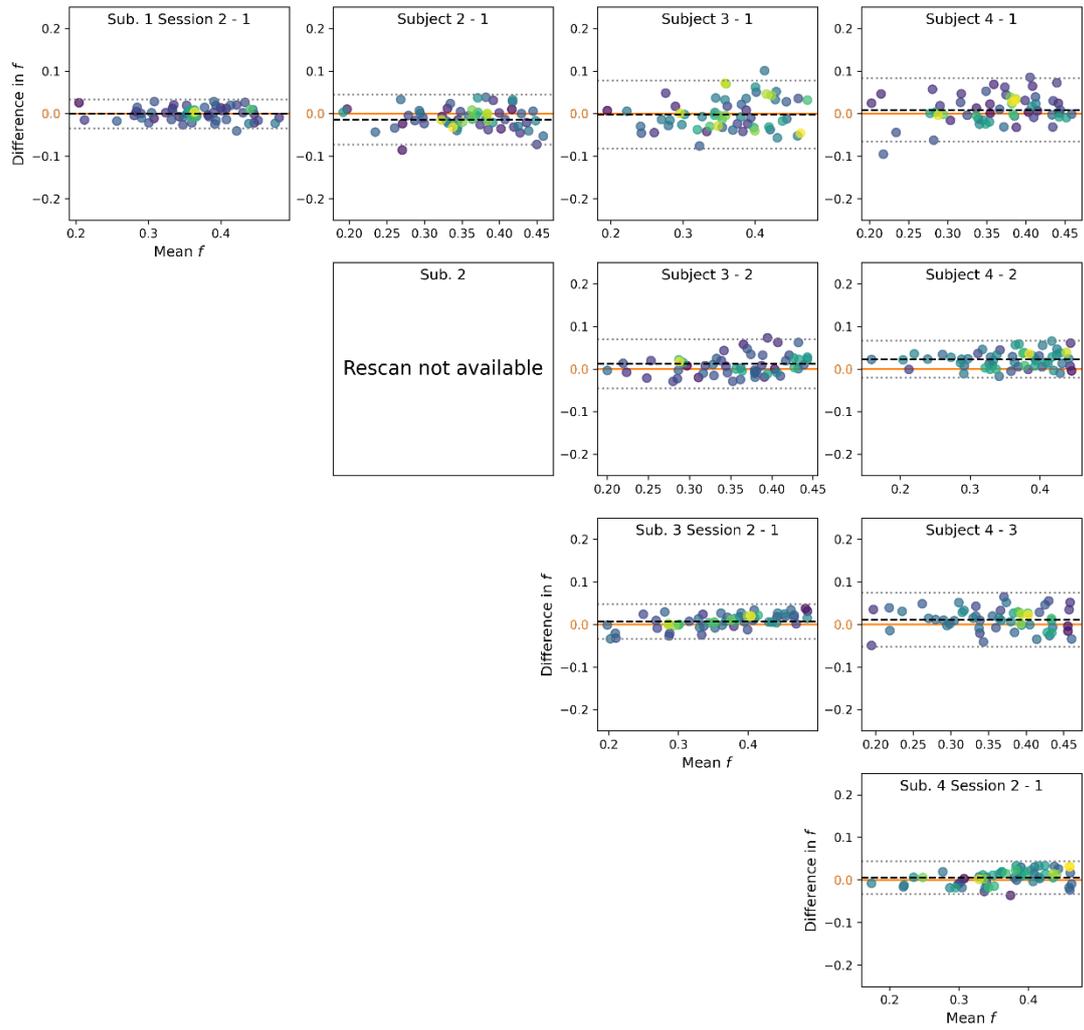

C. Bland-Altman plot of agreement between $f$ median ROI estimations using $NEXI_{Dot}$

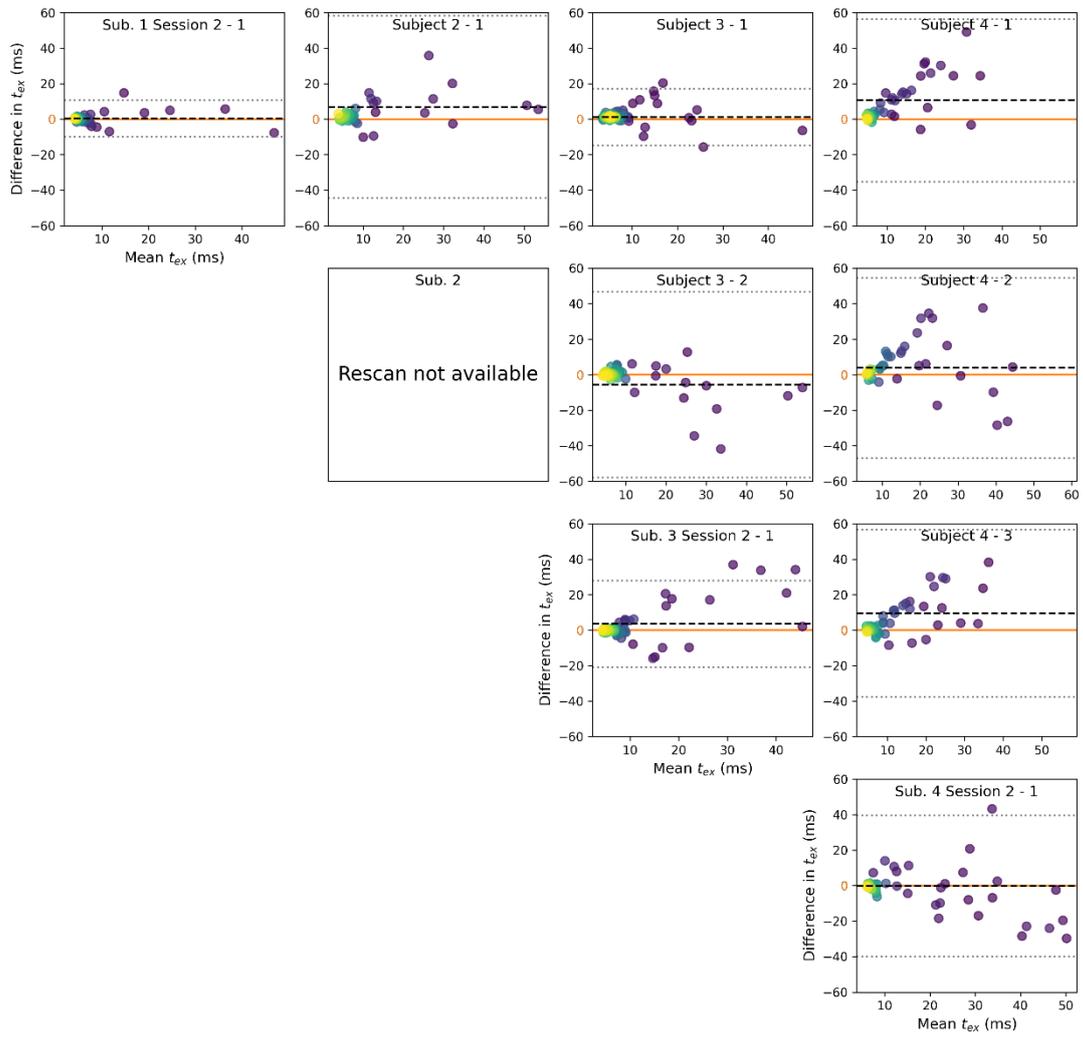

**E.**

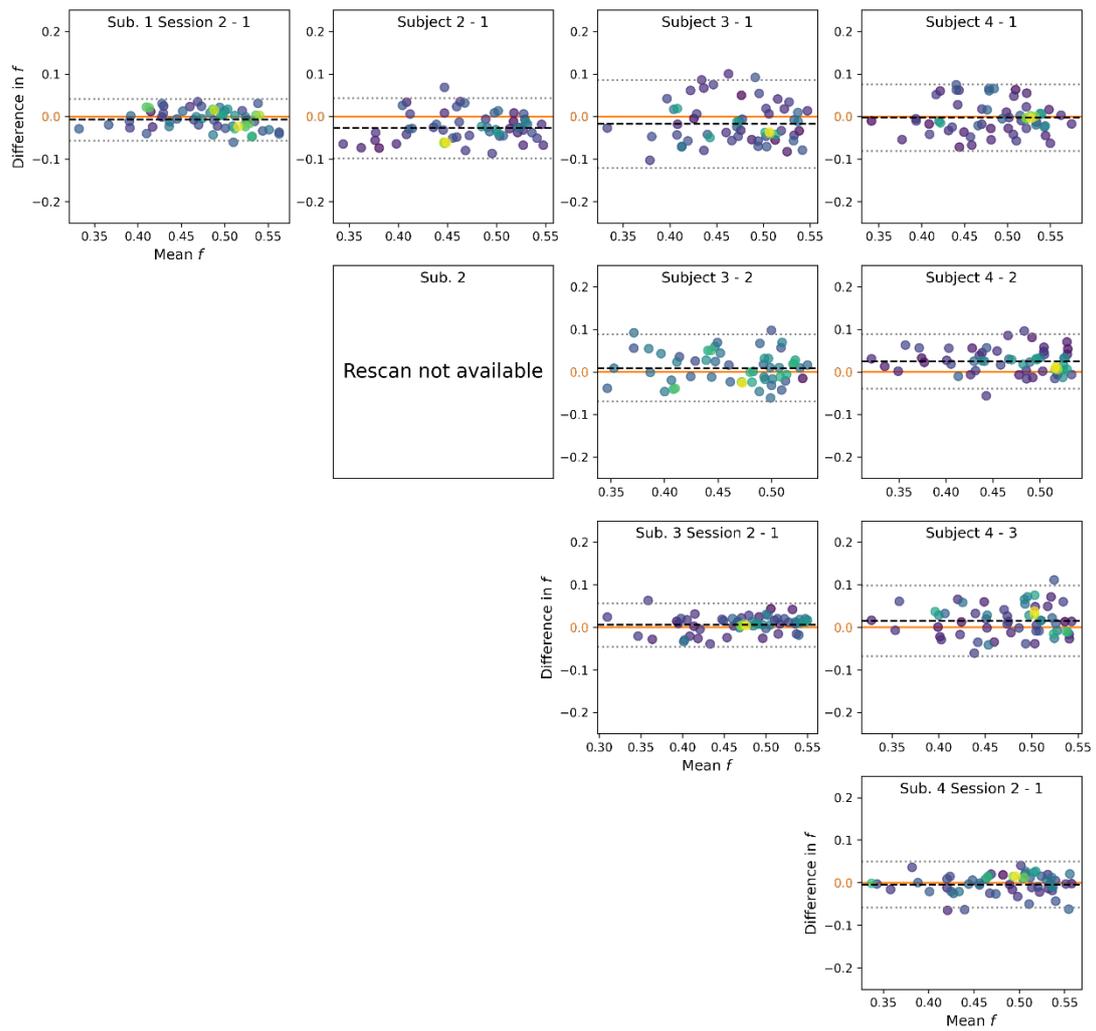

Bland-Altman plot of agreement between $f$ median ROI estimations using $NEXI_{Dot,RM}$

**F.**

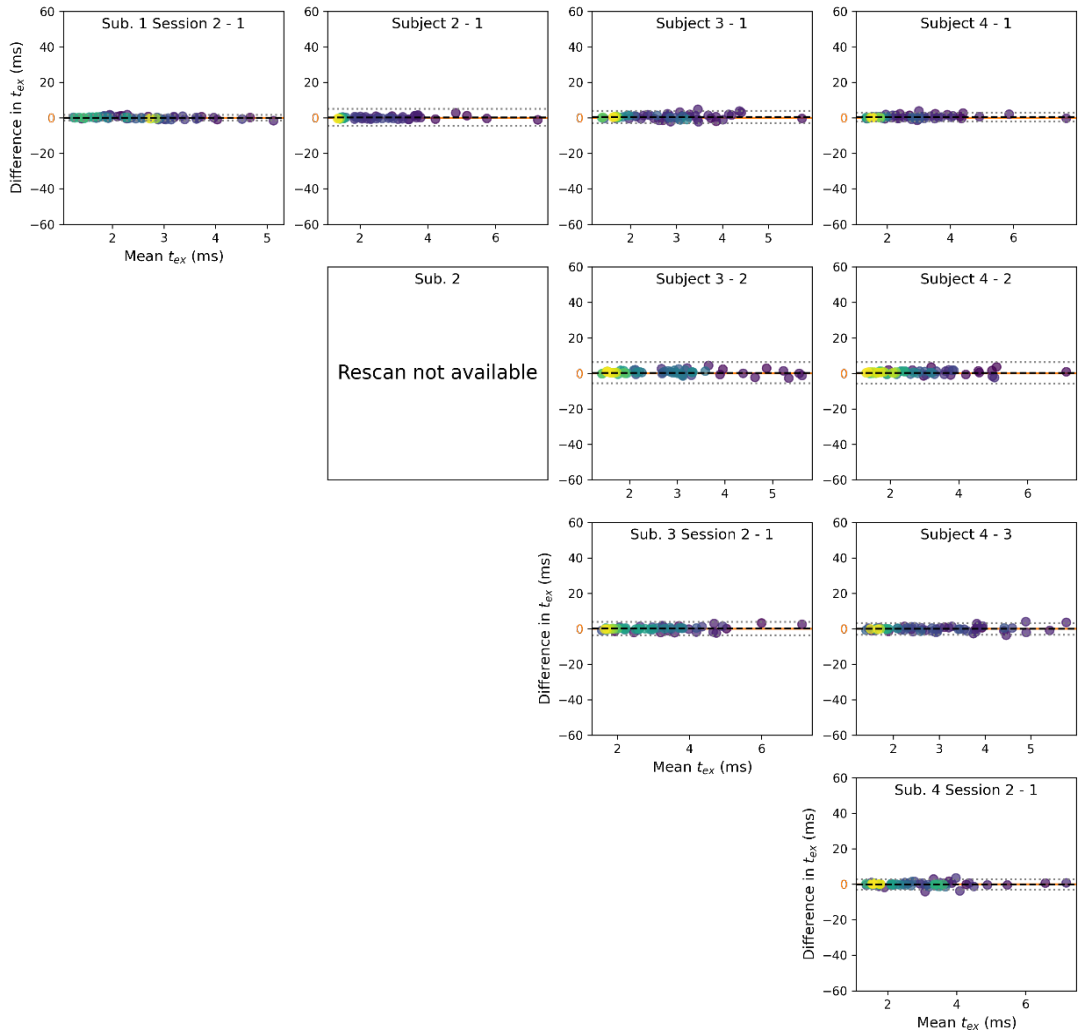

Bland-Altman plot of agreement between $t_{ex}$ median ROI estimations using $NEXI_{Dot,RM}$

**Figure S5.** Bland-Altman plots on the $D_i$ (A) and $D_e$ (B) estimations from the NEXI$_{RM}$ model. Each row and column refer to the same subject. On the diagonal, the two sessions of each subject are compared. In the upper triangle, the results of the first session of each subject are compared.

**A.**

Bland-Altman plot of agreement between $D_i$ median ROI estimations using $NEXI_{RM}$

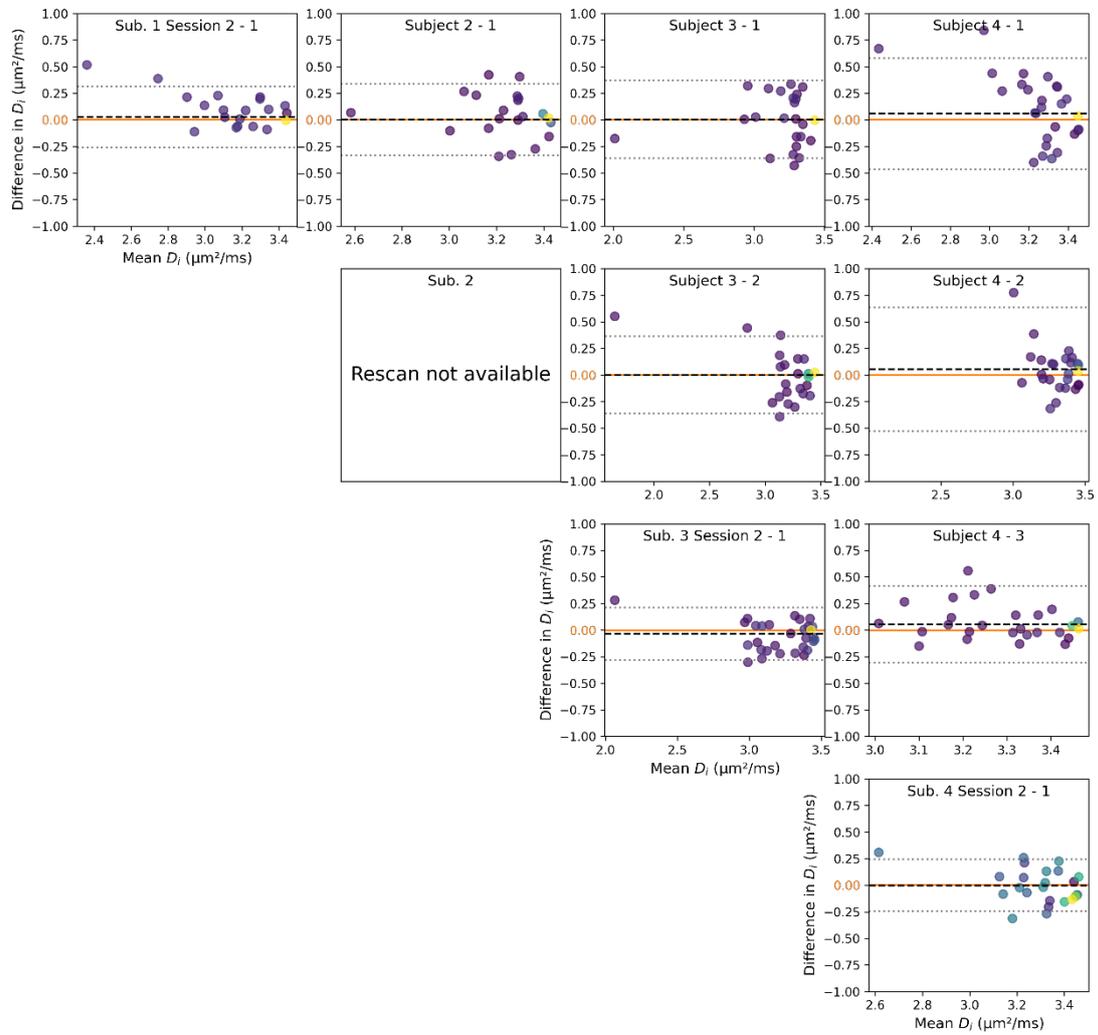

**B.**

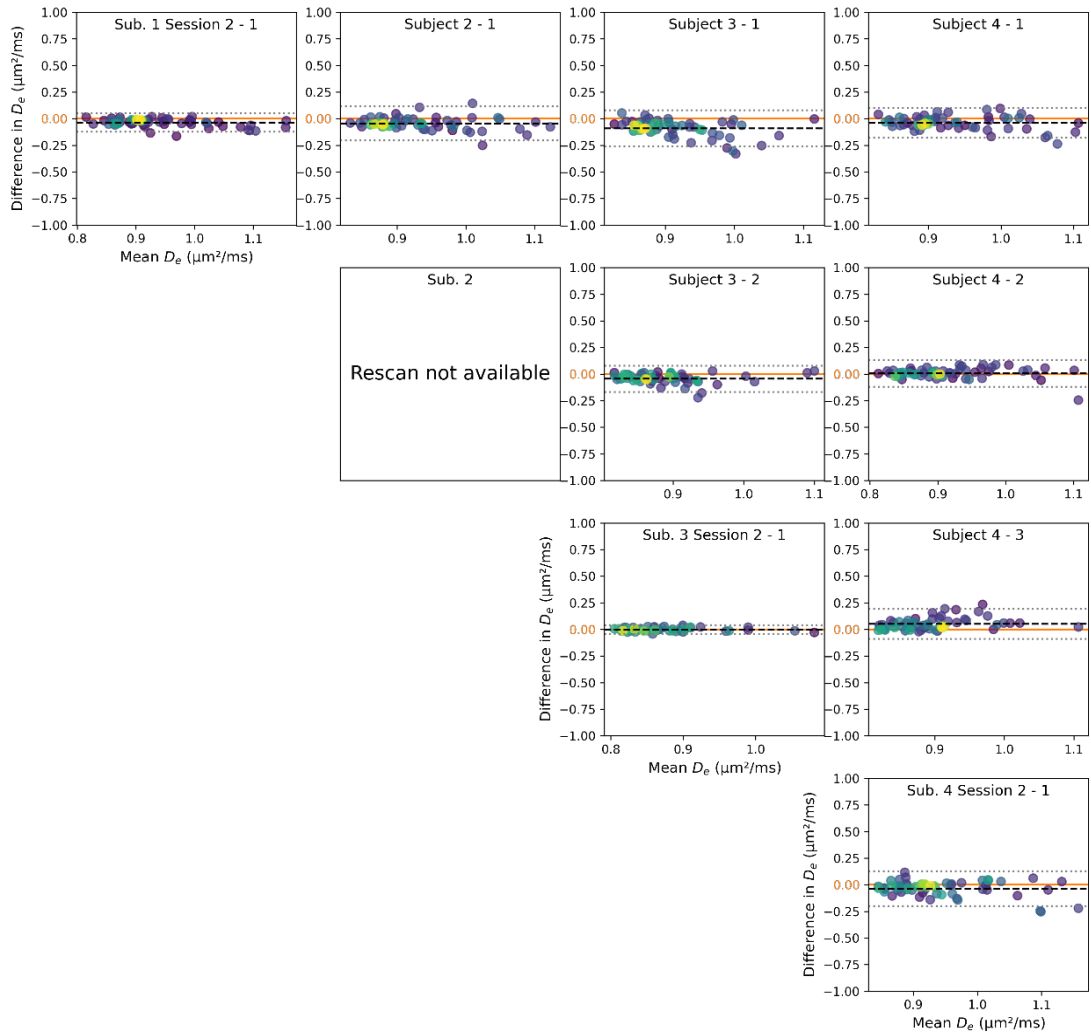

Bland-Altman plot of agreement between $D_e$ median ROI estimations using $NEXI_{RM}$